\documentclass[showpacs,eqsecnum,floatfix,nofootinbib]{revtex4}
\usepackage{bm}
\tighten

\usepackage{amsmath}

\usepackage[dvips]{graphicx,color}

\begin{document}

\hfill{HISKP-TH-0920}
\title{Subtractive renormalization of the NN interaction in chiral
effective theory up to next-to-next-to-leading order: S waves}
\author{C.-J.~Yang$^{1}$, Ch.~Elster$^{1}$, and D.~R.~Phillips$^{1,2}$}
\affiliation{$^1$ Institute of Nuclear and Particle Physics and Department of Physics and
Astronomy,\\ Ohio University, Athens, OH 45701, USA;\\
$^2$ Helmholtz-Institut f\"{u}r Strahlen- und Kernphysik (Theorie), Universit\"{a}t Bonn, D-53115 Bonn, Germany}
\email{cjyang, elster, phillips@phy.ohiou.edu}
\date{\today }

\begin{abstract}
We extend our subtractive-renormalization method in order to evaluate the $^1
$S$_0$ and $^3$S$_1$-$^3$D$_1$ NN scattering phase shifts up to
next-to-next-to-leading order (NNLO) in chiral effective theory. We show
that, if energy-dependent contact terms are employed in the NN potential,
the $^1 $S$_0$ phase shift can be obtained by carrying out two subtractions
on the Lippmann-Schwinger equation. These subtractions use knowledge of the
the scattering length and the $^1$S$_0$ phase shift at a specific energy to
eliminate the low-energy constants in the contact interaction from the
scattering equation. For the J=1 coupled channel, a similar renormalization
can be achieved by three subtractions that employ knowledge of the $^3$S$_1$
scattering length, the $^3$S$_1$ phase shift at a specific energy and the $^3
$S$_1$-$^3$D$_1$ generalized scattering length. In both channels a similar
method can be applied to a potential with momentum-dependent contact terms,
except that in that case one of the subtractions must be replaced by a fit
to one piece of experimental data.

This method allows the use of arbitrarily high cutoffs in the
Lippmann-Schwinger equation. We examine the NNLO S-wave phase shifts for
cutoffs as large as 19 GeV and show that the presence of linear energy
dependence in the NN potential creates spurious poles in the scattering
amplitude. In consequence the results are in conflict with empirical data
over appreciable portions of the considered cutoff range. We also identify
problems with the use of cutoffs greater than 1 GeV when momentum-dependent
contact interactions are employed. These problems are ameliorated, but not
eliminated, by the use of spectral-function regularization for the two-pion
exchange part of the NN potential
\end{abstract}

\maketitle

\vspace{10mm}

\section{Introduction}

\bigskip Chiral perturbation theory ($\chi $PT) has now been applied to the
problem of nucleon-nucleon (NN) interactions for almost two decades. Since
Refs.~\cite{We90,We91}, chiral potentials have been deduced from $\chi$PT
Lagrangians~\cite{Or96, Ka97, Ep99, EM02,EM03, Ep05} through the computation
of all NN-irreducible diagrams that occur up to a given order in $\chi$PT.
The resulting potential includes a variety of contact interactions, which
parameterize our ignorance of physics coming from energies higher than the
chiral-symmetry-breaking scale, $\Lambda _{\chi}$. Some of these contact
interactions occur in the $\pi$N Lagrangian, and the low-energy constants
(LECs) that multiply these operators can be determined independently from
the pion-nucleon scattering data~\cite{Fe98, Bu00}. More of a
challenge---both conceptual and practical---is posed by the string of
operators that represent the short-distance physics in the NN sector. These
operators renormalize the divergent loops found when calculating the $\chi$%
PT potential, and so encode the contribution to observables from high-energy
NN states. Each such contact interaction is associated with an unknown LEC. $%
\chi$PT power counting applied to the NN potential mandates that at a fixed
order in the chiral expansion only a finite number of LECs enter the
potential. For instance, at $O(P^3)$ in the NN S-waves (which will be where
the bulk of our attention is in this work) there are five contact operators
that must be considered: two act in the ${}^1$S$_0$ channel, two in the ${}^3
$S$_1$ channel, and one causes coupling between the ${}^3$S$_1$ and ${}^3$D$%
_1$ partial waves.

For low partial waves the $\chi$PT potential is non-perturbative, and the
standard approach involves use of the Lippmann-Schwinger equation (LSE) to
reconstruct the full NN amplitude from the NN-irreducible part~\cite%
{Le97,Ge99,EM06,Dj07}. (C.f. Refs.~\cite%
{KSW98A,KSW98B,FMS99A,FMS99B,Be02,Ol03,BM04,BB03,LvK08,Be08} where parts of
the potential are treated in perturbation theory.) Since the NN interactions
obtained in $\chi$PT do not fall off as the nucleon momenta $\mathbf{p}$  go
to infinity, a cutoff, denoted here by $\Lambda$, must be imposed on the
intermediate states in the LSE. The LECs are then fitted to data for a
variety of cutoffs, and should absorb any strong (power-law with power $>0$
or logarithmic) dependence of observables on $\Lambda$. A corollary is that
the predictions of the effective theory should not depend on the particular
quantity chosen to fix the LECs, as long as the kinematic point is within
the domain of validity of the chiral effective theory. A potential
which does not have these desirable properties (approximate cutoff and
renormalization-point independence) is not \textquotedblleft properly
renormalized\textquotedblright. If the set of contact operators employed
cannot achieve this we conclude that this ``chiral effective
theory'' ($\chi$ET) is unable to give reliable results, since its
predictions are too sensitive to the treatment of the unknown short-distance
physics at scale $\Lambda$. In Refs.~\cite{Or96,Ep99,EM02,EM03,Ep05} the
potential $V$ was computed to a fixed order, and then the NN LECs that
appear in $V$ were fitted to NN data for a range of cutoffs between 500 and
800~MeV. The resulting predictions---especially the ones obtained with the $%
O(P^4)$ potential derived in Refs.~\cite{EM03,Ep05}---are approximately $\Lambda-$independent and describe NN data with
considerable accuracy.

However, the range of $\Lambda$ considered in these papers is quite narrow.
Does the fact that these analyses only consider $\Lambda \leq 800$~MeV
represent an intrinsic limitation on the $\chi$ET approach?

In the case of leading-order potential (which is $O(P^0)$ and whose
long-range part consists only of one-pion exchange) the answer has been
shown to be ``No" in the $^1$S$_0$ and and $^{3}$S$_{1}-^{3}$D$_{1}$
channels. The problem can be properly renormalized for arbitrarily large $%
\Lambda$'s with only constant contact interactions present in $V$ in each
channel---in accordance with the $\chi$PT power counting~\cite{SDW94, Fr99,
ES01, Be02, PVRA04B, PVRA05, NTvK05, Bi06, Ya08}~\footnote{%
In the ${}^1$S$_0$ channel the constant must be $m_q$-dependent~\cite{KSW96}
in order to absorb a logarithmic divergence. This violates the power
counting, but is of no practical consequence as regards the description of
laboratory NN data.}. However, once $\Lambda > 600$ MeV the $\chi$PT
counting for the NN potential fails in waves with $L > 0$ where the tensor
force from one-pion exchange is attractive~\cite%
{ES01,NTvK05,Bi06,PVRA06B,Ya09}. That counting predicts no contact
interaction in these channels, but a contact interaction is necessary to
stabilize the LSE predictions at these cutoffs. Furthermore, in these
partial waves it has been found that even when higher-order $\chi$PT
potentials are considered and the phase shifts obtained from the LSE are
approximately cutoff independent, the results can still show significant
renormalization-point dependence~\cite{Ya09}.

At next-to-leading order (NLO, $O(P^2)$) and next-to-next-to-leading order
(NNLO, $O(P^3)$), the $\chi$PT potential includes diagrams with
two-pion-exchange (TPE). Since the loop integrals there diverge as the
square or cube of the momentum transfer $|\mathbf{q}| \equiv |\mathbf{p}%
^{\prime}-\mathbf{p}|$, the TPE must be associated with contact terms up to
order $P^2$---just as one would have predicted using naive-dimensional
analysis in powers of NN momentum.

In Refs.~\cite{Ep99,epsfr,Ti05,En08} the role of various different $O(P^2)$
contact terms in the LSE was examined and the renormalization of the NN
S-waves was discussed. Refs.~\cite{Ep99,epsfr,En08} found that
momentum-dependent contact terms had difficulty absorbing the cutoff
dependence at larger cutoffs, and also that the contribution from the
(supposedly higher-order) TPE was larger than one-pion exchange. (See Ref.~%
\cite{Sh08} for a similar conclusion in the ${}^1$S$_0$ channel using a
co-ordinate-space analysis.) Ref.~\cite{En08} also considered an
energy-dependent $O(P^2)$ contact interaction in the ${}^1$S$_0$ channel and
showed that the use of a contact interaction linear in energy (which
formally should be equivalent to the momentum-dependent interaction at the
orders in the chiral expansion being considered here) leads to the
appearance of resonances there for $\Lambda \rightarrow \infty$. In either
case, the results of these studies suggest that for cutoffs larger than
about 1 GeV the contribution from contact terms dominates over that from the
long-range part of the potential, raising the question of whether $\chi$PT
is acting as a phenomenological fit form, rather than an effective theory.
As argued in Refs.~\cite{EM06,Ya09,Sh08}, the chiral-symmetry-breaking scale
could be responsible for this situation.

This suggests that there may be a critical cutoff $%
\sim \Lambda _{\chi }$ above which it does not make sense to iterate the TPE
inside the LSE\footnote{See \cite{EG09} for a recent analytic approach which yields a similar conclusion.}.
It is the purpose of this
paper to critically examine the behavior of the NN phase shifts predicted by 
$\chi$ET at NLO and NNLO in the ${}^1$S$_0$ and ${}^3$S$_1$--${}^3$D$_1$
channels over a range of cutoffs from $\Lambda=500$ MeV to $\Lambda > 2$
GeV. In this way we hope to identify this critical cutoff, if it exists, and
discuss the signatures of, and mechanisms responsible for, its appearance.

In order to address these issues we extend our previously developed
subtractive-renormalization method~\cite{Ya08,Ya09} to evaluate the $^{1}$S$%
_{0}$ and $^{3}$S$_{1}-^{3}$D$_{1}$ phase shifts with the higher-order $\chi$%
ET potentials. If energy-dependent contact interactions are employed in
these potentials we show how to relate the $\chi$ET phase shifts to on-shell
quantities. For momentum-dependent contact interactions we formulate the
scattering equations in terms of two (one) on-shell quantities in the
triplet (singlet) channel and then fit the single remaining unknown
parameter to experimental data. This allows us to easily go to high cutoffs
in the LSE and provides clean information about the effect of the
renormalization point. Our method works for any long-range potential. Here
we consider long-range parts evaluated using both dimensional regularization
(DR) and spectral-function-regularization (SFR)\cite{epsfr} at $O(P^2)$ and $%
O(P^3)$. The forms of these potentials were given in Ref.~\cite{Ya09}. Since
issues like differing extractions of the $\pi$N LECs $c_1$, $c_3$ and $c_4$
could produce sizeable changes in the phase shifts we are not overly focused
on the quality of our fit. Instead, our purpose is to see whether these
potentials behave self-consistently with respect to cutoffs in the LSE.
Thus, our strategy is to evaluate in detail the cutoff-dependence of the
S-wave phase shifts, and examine the renormalization-point dependence. These
features depend much less on the precise choice made for $c_1$, $c_3$ and $%
c_4$.

The structure of our work is as follows. First, in Sec.~\ref{sec-singlet},
we introduce our subtractive-renormalization method and explain how to deal
with both momentum- and energy-dependent contact terms in the singlet
channel. We take as input to the method the ${}^1$S$_0$ scattering length, $%
a_s$, and the phase shift at a particular energy, $E^*$. Then, in Sec.~\ref%
{sec-triplet}, we extend our subtractive-renormalization method to the
coupled-channels case and deal with constant, energy-dependent and
momentum-dependent contact terms. While the constant contact term is, as in
Ref.~\cite{Ya08}, solved by one subtraction with $a_t$ as input, for the
energy-dependent contact term we pin down the three unknown constants with
information with $a_t$, the generalized scattering length $\alpha_{20}$ that
governs the low-energy behavior of $\epsilon_1$, and the phase shift at a
particular energy as inputs. In Sec.~\ref{sec-result1}, we discuss our
results in the $^1$S$_0$ channel and their implications. In Sec.~\ref%
{sec-result2}, we discuss our results in the $^3$S$_1$-${}^3$D$_1$ channel
and their implications. We summarize our findings in Sec.~\ref{sec-con}.

\section{Subtractive renormalization in the singlet channel}

\label{sec-singlet} The short-range part of the potential, i.e., the contact
terms, is a parameterization of unknown high-energy physics. Before
introducing our subtractive renormalization scheme, we need to decide what
types of contact term we should consider. Throughout this work we will
consider chiral two-pion-exchange~\cite{Or96,Ka97,Re99,Ep99} as the
long-range part of the NN potential. This is associated with contact terms
up to $O(P^2)$. However, since for non-relativistic particles we have $%
\mathbf{p}^2/M \sim E$ there is the possibility to consider
energy-dependent, instead of momentum-dependent contact interactions. In
particular, this may have certain advantages in terms of evading theorems
that limit the impact of short-distance potentials on phase shifts~\cite%
{Wi55,PC96}. Thus, in the ${}^1$S$_0$ channel we will consider the following
three types of contact terms:

\begin{itemize}
\item[(A)] $v_{SR,0}=\lambda $

\item[(B)] $v_{SR,0}(E)=\lambda +\gamma E$

\item[(C)] $v_{SR,0}(p^{\prime},p)=\lambda +C_2 \ (p^{2}+p^{\prime 2})$,
\end{itemize}

where $\lambda ,$ $\gamma ,$ $C_2$ are unknown constants. (Of course, the
numerical value of $\lambda $ is different for each case.) The overall
potential is then given by 
\begin{equation}
v_0(p^{\prime},p;E)=v_{SR,0}(p^{\prime},p;E) + v_{LR,0}(p^{\prime},p;E),
\end{equation}
with the subscripts $_{SR}$ and $_{LR}$ standing for the short- and
long-range parts of the potential. Specifically, in the ${}^1$S$_0$ channel
we have 
\begin{equation}
v_{LR,0}=\langle 000|V_C + W_C - 3 [V_S + W_S + \mathbf{q}^2 (V_T +
W_T)]|000 \rangle,
\end{equation}
with expressions for the S-wave projected central, spin, and tensor
isoscalar and isovector potentials taken from Ref.~\cite{Ka97,epsfr} and
collected in Ref.~\cite{Ya09}.

\subsection{The constant contact interaction}

The constant contact term can be evaluated by one subtraction with the ${}^1$%
S$_0$ scattering length $a_s$ as input~\cite{Ya08}. This is possible because
knowledge of the on-shell amplitude plus the long-range potential yields the
fully off-shell amplitude, as long as only a constant contact interaction is
present~\cite{HM01,AP04,Ya08}. The step from the on-shell amplitude at an
arbitrary energy $E^*$, $t(p_0^*,p_0^*;E^*)$ (with $p_0^* \equiv ME^*$), to
the half-off-shell amplitude at the same energy is explained in Appendix~\ref%
{appendixa}. The symmetry property of the amplitude $%
t(p,p_0^*;E^*)=t(p_0^*,p;E^*)$ then means that the same sequence of
manipulations can be used to construct $t(p^{\prime},p;E^*)$. With the
fully-off-shell amplitude at a single energy in hand, it is well known how
to construct the $t$-matrix at any arbitrary energy~\cite{Fr99}. In operator
form we have 
\begin{equation}
t(E)=t(E^*) + t(E^*)[g_0(E) - g_0(E^*)] t(E),  \label{eq:EEstar}
\end{equation}
where $g_0$ is the free resolvent of the Lippmann-Schwinger equation, 
\begin{equation}
g_0(E;p)=\frac{1}{E^+ - p^2/M}.
\end{equation}
(Here $M$ is the nucleon mass and $E^+=E+i \varepsilon$, with $\varepsilon$
a positive infinitesimal.)

\subsection{Energy-dependent contact interaction}

\label{sec-1S0endep}

In this case there are two unknown constants. The strategy is to use one
subtraction (as in Ref.~\cite{Ya08}) to first eliminate the constant part of
the contact term. For the energy-dependent contact term we then perform a
second subtraction to eliminate the remaining unknown.

We start from the partial-wave LS equation, which is written explicitly in
this channel as 
\begin{eqnarray}
t_0(p^{\prime },p;E)=v_{0}(p^{\prime },p;E)+ \frac{2}{\pi }%
M\int_{0}^{\Lambda }\frac{dp^{\prime \prime }\;p^{\prime \prime
}{}^{2}\;v_0(p^{\prime },p^{\prime \prime };E)\;t_{0}(p^{\prime \prime },p;E)%
}{p_{0}^{2}+i\varepsilon -p^{\prime \prime }{}^{2}}.  \label{eq:1a}
\end{eqnarray}
The energy-dependent ${}^1$S$_0$ potential is given by 
\begin{equation}
v_{0}(p^{\prime },p;E)=v_{LR }(p^{\prime},p)+\lambda + \gamma E.
\label{eq:4.3}
\end{equation}
Here $v_{LR}$ represents the long range part of the $\chi$PT  potential,
which is computed up to NLO or NNLO. However, the derivation presented below
holds for any energy-independent long-range potential which is a function of 
$p$ and $p^{\prime }$ and satisfies $v_{LR}(p^{\prime
},p)=v_{LR}(p,p^{\prime })$. From now on we drop the partial-wave index,
since this section only describes the singlet channel $^1$S$_0$.

To further simplify the presentation, we adopt the following operator
notation of the LS equation 
\begin{eqnarray}
t(E)=\lambda +\gamma E+v_{LR } + \left[\lambda +\gamma E+v_{LR } \right] \
g_0(E) \ t(E).  \label{eq:4.4}
\end{eqnarray}
Setting $E=0$ in Eq.~(\ref{eq:4.4}) leads to 
\begin{eqnarray}
t(0)=\lambda +v_{LR } + \left[ \lambda +v_{LR} \right] \ g_0(0) \ t(0).
\label{eq:4.5}
\end{eqnarray}
Equation (\ref{eq:4.5}) contains only one unknown, $\lambda $. Therefore,
the matrix element $t(p^{\prime },p;0)$ can be obtained from one
experimental point, i.e., the NN scattering length, $a_{s}$, using the
procedure explained in Subsection A, and given in detail in Ref.~\cite{Ya08}%
. Using properties of the LSE, Eq.~(\ref{eq:4.5}) can be rewritten as 
\begin{eqnarray}
t(0)&=&\lambda +v_{LR }+t(0)g_0(0) \left(\lambda +v_{LR }\right) \cr &=&%
\left[1+t(0)g_0(0)\right] \ \left[\lambda +v_{LR }\right].  \label{eq:4.65}
\end{eqnarray}
This allows to reconstruct $\lambda$ if we wish to do so: 
\begin{eqnarray}
\lambda +v_{LR }= \left[1+t(0)g_0(0)\right]^{-1} \ t(0).
\label{eq:4.7}
\end{eqnarray}

\noindent Leaving $E$ finite, Eq.~(\ref{eq:4.4}) can be written as 
\begin{eqnarray}
t(E)=\left[ \lambda +\gamma E+v_{LR } \right] \ \left[1+g_0(E)t(E)\right],
\label{eq:4.55}
\end{eqnarray}
leading to a formal expression for the potential term 
\begin{eqnarray}
\lambda +\gamma E+v_{LR }=t(E) \ \left[1+g_0(E)t(E) \right]^{-1}.
\label{eq:4.6}
\end{eqnarray}
Subtracting Eq.(\ref{eq:4.7}) from Eq.(\ref{eq:4.6}) isolates the term $%
\gamma E$, 
\begin{eqnarray}
\gamma E=t(E) \ \left[1+g_0(E)t(E)\right]^{-1}-\left[1+t(0)g_0(0)\right]%
^{-1} \ t(0).  \label{eq:4.75}
\end{eqnarray}
Proceeding in a similar fashion as in Ref.~\cite{Ya08}, we can obtain $%
t(E^{\ast })$ at a fixed energy $E^{\ast }$ from the value of the NN phase
shift at $E^\ast $ (for details see Appendix \ref{appendixa}). We find~%
\footnote{%
Note that some care is required if the standard subtraction method is used
to deal with the singularity when the integration momentum becomes equal to $%
p_0$. In that case it is useful to introduce an additional term in the
integral equation that involves the potential evaluated at the point $%
p^{\prime}=p_0$, $p=p_0$. However, if other techniques are used (e.g.
contour rotation) this additional term is not necessary and so we do not
include it in our derivation.} 
\begin{eqnarray}
\lambda +\gamma E^{\ast }+v_{LR }= t(E^{\ast }) \ \left[1+g_0(E^{\ast
})t(E^{\ast })\right]^{-1}=[1+t(E^*)g_0(E^*)]^{-1} t(E^*).  \label{eq:4.8}
\end{eqnarray}
Taking the difference of Eqs.~(\ref{eq:4.8}) and (\ref{eq:4.6}) leads to 
\begin{eqnarray}
\gamma (E^{\ast }-E) = \left[1+t(E^{\ast }) g_0(E^{\ast })\right]^{-1}
t(E^\ast ) - t(E) \ \left[ 1+g_0(E)t(E) \right]^{-1}.  \label{eq:4.9}
\end{eqnarray}
To eliminate $\gamma $, we insert Eq.~(\ref{eq:4.75}) into Eq.~(\ref{eq:4.9}%
), 
\begin{eqnarray}
\Big(t(E)[1+g_0(E)t(E)]^{-1} &-& [1+t(0)g_0(0)]^{-1}t(0) \Big) \left( \frac{%
E^\ast}{E}-1 \right) \cr & =& \left[1+t(E^{\ast }) g_0(E^{\ast })\right]%
^{-1} t(E^\ast ) - t(E) \ \left[ 1+g_0(E)t(E) \right]^{-1}.  \label{eq:4.95}
\end{eqnarray}
Rearranging Eq.~(\ref{eq:4.95}) and multiplying both sides with $1+g_0(E)t(E)
$ from the right and $1+t(0)g_0(0)$ from the left, we arrive at 
\begin{eqnarray}
t(E) + t(0) [g_0(0) -g_0(E)] t(E) +\frac{E}{E^*} \Big\{ t(0) - \left[%
1+t(0)g_0(0)\right] \alpha t(E^{\ast })\Big\} \ g_0(E) t(E) \cr = \left(1-%
\frac{E}{E^*} \right) t(0) + \frac{E}{E^*} \Big[1+t(0)g_0(0)\Big]\alpha
t(E^{\ast }),  \label{eq:4.96}
\end{eqnarray}
where $\alpha \equiv [1 + t(E^*) g_0(E^*)]^{-1}$, can be calculated
numerically. Since we already obtained the fully off-shell $t(0)$ and $t(E^*)
$, Eq.~(\ref{eq:4.96}) is an integral equation for $t(E)$, which we can
solve by standard methods.

To summarize, we perform two subtractions to the LS equation to eliminate
the two unknown constants $\lambda $ and $\gamma$. The resulting equation
requires as input the scattering length $a_{s}$ and the phase shift at one
specific energy $E^\ast$. The only restriction on $E^*$ is that it must be
within the domain of validity of our theory. Hence one can test the
consistency of the theory by examining the extent to which results depend
upon the choice of $E^*$.

\subsection{Momentum-dependent contact interaction}

The coefficient of the momentum-dependent contact interaction, denoted here by $C_2$, is not
straightforwardly related to any S-matrix element. Thus we cannot apply our
subtraction procedure in the case of a momentum-dependent contact
interaction. Hence we instead adopt a ``mixed" procedure, which involves a
single subtraction plus fitting of $C_2$.~\footnote{%
Note that this is different from the subtractive procedure of Ref.~\cite%
{Ti05} for dealing with the same contact interaction. There the Born
approximation is assumed to be valid for large negative energies.}

This ``mixed" procedure is carried out as follows: First we guess a value
for the constant $C_2$, and then insert it into the half-off-shell and
on-shell LSE: 
\begin{eqnarray}  \label{eq:1}
t(p,0;0) &=&v_{LR}(p,0)+\lambda +C_2 p^{2} - \frac{2}{\pi }M
\int_{0}^{\Lambda }dp^{\prime }\;\left[ v_{LR}(p,p^{\prime })+\lambda +C_2
(p^{2}+p^{\prime 2}) \right] \;t(p^{\prime },0;0),  \notag \\
\\
t(0,0;0) &=&v_{LR}(0,0)+\lambda -\frac{2}{\pi }M\int_{0}^{\Lambda
}dp^{\prime }\; \left( v_{LR}(0,p^{\prime })+\lambda +C_2 p^{\prime 2}
\right) \;t(p^{\prime },0;0).  \label{eq:2}
\end{eqnarray}
Taking the difference of the two equations cancels the constant $\lambda$
and leads to 
\begin{eqnarray}
t(p,0;0)&=&v_{LR}(p,0)-v_{LR}(0,0)+\frac{a_{s}}{M}+C_2 p^{2}  \notag \\
&-&\frac{2}{\pi } M\;\int_{0}^{\Lambda }dp^{\prime }\; \;\left(
v_{LR}(p,p^{\prime })-v_{LR}(0,p^{\prime })+C_2 p^{2} \right) \;t(p^{\prime
},0;0).  \label{eq:3}
\end{eqnarray}
Using the already-chosen value of $C_2$, together with the experimental
value of $a_s$, we can solve for $t(p^{\prime},0;0)$ from Eq.~(\ref{eq:3}).
Then there is a consistency condition that determines the value of $\lambda$%
. This equation can be easily derived from Eq.~(\ref{eq:2}):%
\begin{eqnarray}
\lambda =\frac{\frac{a_{s}}{M}-v_{LR}(0,0) + \frac{2}{\pi }M\;
\int_{0}^{\Lambda }dp^{\prime }\;\left( v_{LR}(0,p^{\prime })+C_2 p^{\prime
2} \right) \;t(p^{\prime },0;0)}{1-\frac{2}{\pi } M\;\int_{0}^{\Lambda
}dp^{\prime }\;t(p^{\prime },0;0)}.  \label{eq:4}
\end{eqnarray}
The above equation gives the value of $\lambda$ that is consistent with the
experimental value of the scattering length and our choice of $C_2$. It thus
defines a relationship $\lambda=\lambda(C_2;a_s)$. (Note that---in spite of
the form of Eq.~(\ref{eq:4})---the relationship is not linear, since $%
t(p^{\prime},0;0)$ is also affected by the choice of $C_2$.) Therefore, when
trying to determine $\lambda$ and $C_2$ we only need to guess $C_2$ and can
then obtain $\lambda$ from Eqs.~(\ref{eq:3}) and (\ref{eq:4}). These two
constants are then entered into the on-shell LS equation which is solved for
the phase shifts. Finally $C_2$ is adjusted to fit the desired observable.
In Sec.~\ref{sec-result1} we will examine the results obtained when $C_2$ is
adjusted to reproduce the ${}^1$S$_0$ effective range, $r_0$, and those
found when we enforce the requirement that the theory correctly predict the
phase shift at a particular energy $E^\ast$.

\section{Subtractive renormalization in the J=1 triplet channel}

\label{sec-triplet}

Since the NN interaction is non-central, the $S=1$ waves constitute a
coupled-channel problem. Here we consider the ${}^3$S$_1$-${}^3$D$_1$
coupled partial waves, and we again consider three different contact terms:

\begin{itemize}
\item[(A)] $\left( 
\begin{array}{cc}
\lambda & 0 \\ 
0 & 0%
\end{array}
\right) $

\item[(B)] $\left( 
\begin{array}{cc}
\lambda +C_2 (p^{2}+p^{\prime 2}) & \lambda_{t} \ p^{\prime 2} \\ 
\lambda _{t} \ p^{2} & 0%
\end{array}
\right) $

\item[(C)] $\left( 
\begin{array}{cc}
\lambda +\gamma E & \lambda_t \ p^{\prime 2} \\ 
\lambda_t \ p^{2} & 0%
\end{array}
\right). $
\end{itemize}

Here we write the contact terms explicitly in their matrix form, where the
diagonal represents the direct channels $^3$S$_1$-$^3$S$_1$ and $^3$D$_1$-$^3
$D$_1$ and the off-diagonal the channels $^3$S$_1$-$^3$D$_1$ and $^3$D$_1$-$%
^3$S$_1$. The unknown constants are $\lambda $, $\gamma$ (or $C_2$) and $%
\lambda _{t}$, and their value is different in each case. Case A is the
leading-order contact interaction discussed in Refs.~\cite{Be02,Ya08}. Cases
B and C include the structures that appear at NLO [$O(P^2)$] in the standard
chiral counting for short-distance operators.

At this order in the chiral expansion the J=1 coupled-channel problem
acquires two additional contact interactions. One is of the form $%
\boldsymbol{\sigma}_1 \cdot \mathbf{q} \ \boldsymbol{\sigma}_2 \cdot \mathbf{%
q}$~\cite{Ep99}, and hence gives a non-zero matrix element for the $^3$S$_1$-$^3$%
D$_1$ transition. The other can be written as either an energy-dependent or
momentum-dependent piece of the $^3$S$_1$-$^3$S$_1$ potential, although
until now most works on $\chi$ET have considered only the latter~\cite%
{Or96,Ep99,EM03,Ep05}, but see Ref.~\cite{En08}. For the energy-dependent
contact interaction we will show below that three subtractions can be
performed to eliminate the unknown LECs. This leaves us with an integral
equation for the $t$-matrix at an arbitrary energy that takes as input three
experimental quantities. For case B, due to the momentum-dependence in the
contact term, we use a double-subtraction-plus-one-fit scheme to solve the
LSE. The procedures for solving case B and case C are quite similar, as they both
employ subtractions to eliminate $\lambda $ and $\lambda _{t}$.

\subsection{Constant contact interaction}

This case can be solved with a single subtraction, as described in Ref.~\cite%
{Ya08}. Once again, the key fact is that once $a_t$, the triplet scattering
length, is known, the value of $t(0,0;0)$ is fixed, and that, together with
knowledge of the long-range potential, is sufficient to obtain the two
non-zero elements of $t(p,0;0)$. The procedure by which we obtain these two
elements, $t_{00}$ and $t_{20}$, can be reconstructed from Appendix~\ref%
{appendixb} by setting $C_2=0$ in the first part of the derivation presented
there. Hermiticity of $v$ then implies that $t_{00}(p,0;0)=t_{00}(0,p;0)$
and $t_{20}(p,0;0)=t_{02}(0,p;0)$, and the same set of manipulations can
then be used to obtain $t_{l^{\prime}l}(p^{\prime},p;0)$. The operator
equation (\ref{eq:EEstar}), with $E^*=0$, then allows us to reconstruct $t(E)
$, albeit this time as a two-by-two matrix of operators, rather than the
single operator $t$ that is needed to describe the ${}^1$S$_0$ channel.

\subsection{Momentum-dependent central part}

In this case we will eliminate two constants from the integral equation: $%
\lambda$, and the coefficient of the tensor short-distance interaction $%
\lambda_t$. We will then adjust the coefficient $C_2$ to reproduce one piece
of `experimental' information.

As usual, we begin by writing the on-shell and half-off-shell partial-wave
LS equations: 
\begin{eqnarray}  \label{eq:5}
\left( 
\begin{array}{cc}
t_{00}(p_{0},p_{0};E) & t_{02}(p_{0},p_{0};E) \\ 
t_{20}(p_{0},p_{0};E) & t_{22}(p_{0},p_{0};E)%
\end{array}
\right) & =& \left( 
\begin{array}{cc}
\lambda +C_2 (p_{0}^{2}+p_{0}^{2})+v_{LR }^{00}(p_{0},p_{0}) & \lambda
_{t}p_{0}^{2}+v_{LR }^{02}(p_{0},p_{0}) \\ 
\lambda _{t}p_{0}^{2}+v_{LR }^{20}(p_{0},p_{0}) & v_{LR }^{22}(p_{0},p_{0})%
\end{array}
\right)  \notag \\
+ \frac{2}{\pi }M \int_{0}^\Lambda &dp^{\prime } & \frac{p{}^{\prime 2}}{%
p_{0}^{2}-p^{\prime }{}^{2}+i\varepsilon } \left( 
\begin{array}{cc}
\lambda +C_2 (p_{0}^{2}+p^{\prime 2})+v_{LR }^{00}(p_{0},p^{\prime }) & 
\lambda _{t}p^{\prime 2}+v_{LR }^{02}(p_{0},p^{\prime }) \\ 
\lambda _{t}p_{0}^{2}+v_{LR }^{20}(p_{0},p^{\prime }) & v_{LR
}^{22}(p_{0},p^{\prime })%
\end{array}
\right)  \notag \\
& \times & \left( 
\begin{array}{cc}
t_{00}(p^{\prime },p_{0};E) & t_{02}(p^{\prime },p_{0};E) \\ 
t_{20}(p^{\prime },p_{0};E) & t_{22}(p^{\prime },p_{0};E)%
\end{array}
\right)
\end{eqnarray}

\begin{eqnarray}  \label{eq:6}
\left( 
\begin{array}{cc}
t_{00}(p,p_{0};E) & t_{02}(p,p_{0};E) \\ 
t_{20}(p,p_{0};E) & t_{22}(p,p_{0};E)%
\end{array}
\right) &=&\left( 
\begin{array}{cc}
\lambda +C_2 (p^{2}+p_{0}^{2})+v_{LR }^{00}(p,p_{0}) & \lambda
_{t}p_{0}^{2}+v_{LR }^{02}(p,p_{0}) \\ 
\lambda _{t}p^{2}+v_{LR }^{20}(p,p_{0}) & v_{LR }^{22}(p,p_{0})%
\end{array}
\right)  \notag \\
+\frac{2}{\pi }M\int_{0}^{\Lambda } & dp^{\prime } & \frac{p{}^{\prime 2}}{
p_{0}^{2}-p^{\prime }{}^{2}+i\varepsilon } \left( 
\begin{array}{cc}
\lambda +C_2 (p^{2}+p^{\prime 2})+v_{LR }^{00}(p,p^{\prime }) & \lambda
_{t}p^{\prime 2}+v_{LR }^{02}(p,p^{\prime }) \\ 
\lambda _{t}p^{2}+v_{LR }^{20}(p,p^{\prime }) & v_{LR }^{22}(p,p^{\prime })%
\end{array}
\right)  \notag \\
&\times & \left( 
\begin{array}{cc}
t_{00}(p^{\prime },p_{0};E) & t_{02}(p^{\prime },p_{0};E) \\ 
t_{20}(p^{\prime },p_{0};E) & t_{22}(p^{\prime },p_{0};E)%
\end{array}
\right),
\end{eqnarray}
where $p_{0}^{2}/M=E$. The subscript (superscript) in $t_{l^{\prime}l}
(v^{ll^{\prime}}_{LR})$ indicates the angular-momentum quantum number of the
channels. To simplify notation, we write the $2\times2$ matrix as t.
Subtracting Eq.~(\ref{eq:6}) from Eq.~(\ref{eq:5}) cancels $\lambda $, 
\begin{eqnarray}
\mathbf{t}(p,p_{0};E)&-&\mathbf{t}(p_{0},p_{0};E) =\left( 
\begin{array}{cc}
C_2 (p^{2}-p_{0}^{2})+v_{LR }^{00}(p,p_{0})-v_{LR }^{00}(p_{0},p_{0}) & 
v_{LR }^{02}(p,p_{0})-v_{LR }^{02}(p_{0},p_{0}) \nonumber \\ 
\lambda _{t}(p^{2}-p_{0}^{2})+v_{LR }^{20}(p,p_{0})-v_{LR }^{20}(p_{0},p_{0})
& v_{LR }^{22}(p,p_{0})-v_{LR }^{22}(p_{0},p_{0})%
\end{array}
\right) \\
&+&\frac{2}{\pi }M\int_{0}^{\Lambda } dp^{\prime } \frac{p{}^{\prime 2}} {%
p_{0}^{2}-p^{\prime }{}^{2}+i\varepsilon } \left( 
\begin{array}{cc}
C_2 (p^{2}-p_{0}^{2})+v_{LR }^{00}(p,p^{\prime })-v_{LR
}^{00}(p_{0},p^{\prime }) & v_{LR }^{02}(p,p^{\prime })-v_{LR
}^{02}(p_{0},p^{\prime }) \\ 
\lambda _{t}(p^{2}-p_{0}^{2})+v_{LR }^{20}(p,p^{\prime })-v_{LR
}^{20}(p_{0},p^{\prime }) & v_{LR }^{22}(p,p^{\prime })-v_{LR
}^{22}(p_{0},p^{\prime })%
\end{array}
\right) \mathbf{t}(p^{\prime },p_{0};E).  \notag \\
&&  \label{eq:7}
\end{eqnarray}

\noindent Letting $E\rightarrow 0$, i.e. $p_{0}\rightarrow 0$, in Eq.~(\ref%
{eq:7}) leads to 
\begin{eqnarray}  \label{eq:7.5}
\mathbf{t}(p,0;0)-\mathbf{t}(0,0;0)&=&\left( 
\begin{array}{cc}
C_2 p^{2}+v_{LR }^{00}(p,0)-v_{LR }^{00}(0,0) & 0 \\ 
\lambda _{t}p^{2}+v_{LR }^{20}(p,0) & 0%
\end{array}
\right)  \notag \\
-\frac{2}{\pi }M\int_{0}^{\Lambda }&dp^{\prime }& \left( 
\begin{array}{cc}
C_2 p^{2}+v_{LR }^{00}(p,p^{\prime })-v_{LR }^{00}(0,p^{\prime }) & v_{LR
}^{02}(p,p^{\prime })-v_{LR }^{02}(0,p^{\prime }) \\ 
\lambda _{t}p^{2}+v_{LR }^{20}(p,p^{\prime }) & v_{LR }^{22}(p,p^{\prime })%
\end{array}
\right) t(p^{\prime },0;0).
\end{eqnarray}

\noindent Here we used the threshold behavior of the partial-wave potential, 
$v_{LR}^{l^{\prime }l}(p,k)$ $\sim p^{l^{\prime }}k^{l}$, to infer that,
e.g. $v_{LR }^{20}(0,p)=0$.

Equation (\ref{eq:7.5}) shows that this feature of $v_{LR}$ has as
consequence that $t_{02}(p,0;0)=t_{20}(0,p;0)=0$ and $%
t_{22}(p,0;0)=t_{22}(0,p;0)=0$. Using Eq.~(\ref{eq:7.5}) we can thus obtain 
the first part of the t-matrix equation that needs to be solved, namely  
\begin{eqnarray}  \label{eq:7.6}
t_{00}(p,0;0)-\frac{a_{t}}{M} &=& C_2 p^{2}+v_{LR }^{00}(p,0)-v_{LR
}^{00}(0,0)  \notag \\
-\frac{2}{\pi }M\int_{0}^{\Lambda }&dp^{\prime }& \Big[ \left( C_2
p^{2}+v_{LR }^{00}(p,p^{\prime })-v_{LR }^{00}(0,p^{\prime }) \right)
t_{00}(p^{\prime },0;0) + \left( v_{LR }^{02}(p,p^{\prime}) - v_{LR
}^{02}(0,p^{\prime}) \right) t_{20}(p^{\prime},0;0)\Big],
\end{eqnarray}
where we have used that $t_{00}(0,0;0)=\frac{a_{t}}{M}$. We observe that $%
\lambda_t$ has been eliminated from this equation, but that we do not have a
closed set of equations, since $t_{20}(p^{\prime},0;0)$ appears on the
right-hand side.

To eliminate $\lambda_t$ we set $p=0$ in Eq.~(\ref{eq:7}) 
\begin{eqnarray}
&&\mathbf{t}(0,p_{0};E)-\mathbf{t}(p_{0},p_{0};E)=  \notag \\
&& \qquad \left( 
\begin{array}{cc}
-C_2 p_{0}^{2}+v_{LR }^{00}(0,p_{0})-v_{LR }^{00}(p_{0},p_{0}) & v_{LR
}^{02}(0,p_{0})-v_{LR }^{02}(p_{0},p_{0}) \\ 
-\lambda _{t} p_{0}^{2}-v_{LR }^{20}(p_{0},p_{0}) & -v_{LR
}^{22}(p_{0},p_{0})%
\end{array}
\right)  \notag \\
&& \qquad \qquad +\frac{2}{\pi }M\int_{0}^{\Lambda }dp^{\prime } \frac{%
p{}^{\prime 2}}{p_{0}^{2}-p^{\prime }{}^{2}+i\varepsilon } \left( 
\begin{array}{cc}
-C_2 p_{0}^{2}+v_{LR }^{00}(0,p^{\prime })-v_{LR }^{00}(p_{0},p^{\prime }) & 
v_{LR }^{02}(0,p^{\prime })-v_{LR }^{02}(p_{0},p^{\prime }) \\ 
-\lambda _{t}p_{0}^{2}-v_{LR }^{20}(p_{0},p^{\prime }) & -v_{LR
}^{22}(p_{0},p^{\prime })%
\end{array}%
\right) \mathbf{t}(p^{\prime },p_{0};E)  \notag \\
.  \label{eq:8}
\end{eqnarray}
In order to use the threshold behavior of $t_{20}$ we need to divide  Eq.~(%
\ref{eq:7}) by $p^{2}-p_{0}^{2}$ and Eq.~(\ref{eq:8}) by $p_{0}^{2}$ and
then add both equations. A quick examination shows that $\gamma$ and $%
\lambda_t$ both drop out of the final result---but the final result is
quite lengthy. Hence, to simplify the notation arising from these
manipulations we define the following map 
\begin{equation}
\mathcal{F} \big[O(p,p^{\prime})\big] \equiv \frac{1}{p^2-p_0^2}\Big( %
O(p,p^{\prime}) - O(p_0,p^{\prime}) \Big),  \label{Fmap}
\end{equation}
where the operator $O$ can represent either the $t$-matrix  or a potential
function. With this notation we obtain 
\begin{eqnarray}
&& \mathcal{F}\big[\mathbf{t}(p,p_0;E)\big] - \mathcal{F}\big[\mathbf{t}%
(0,p_0;E)\big]=  \notag \\
&& \qquad \left( 
\begin{array}{cc}
\mathcal{F}\big[v_{LR }^{00}(p,p_{0})\big] - \mathcal{F}\big[v_{LR
}^{00}(0,p_{0})\big] & \mathcal{F}\big[ v_{LR }^{02}(p,p_{0})\big] - 
\mathcal{F}\big[ v_{LR }^{02}(0,p_{0})\big] \\ 
\mathcal{F}\big[v_{LR }^{20}(p,p_{0})\big] - \mathcal{F}\big[ v_{LR
}^{20}(0,p_{0})\big] & \mathcal{F}\big[ v_{LR }^{22} (p,p_{0})\big] - 
\mathcal{F}\big[ v_{LR }^{22} (0,p_{0})\big]%
\end{array}
\right)  \notag \\
&& \qquad \qquad + \frac{2}{\pi }M\int_{0}^{\Lambda } dp^{\prime } \frac{%
p{}^{\prime 2}}{p_{0}^{2}-p^{\prime }{}^{2}+i\varepsilon } \left( 
\begin{array}{cc}
\mathcal{F}\big[v_{LR }^{00}(p,p^{\prime})\big] - \mathcal{F}\big[v_{LR
}^{00}(0,p^{\prime})\big] & \mathcal{F}\big[ v_{LR }^{02}(p,p^{\prime})\big] %
- \mathcal{F}\big[ v_{LR }^{02}(0,p^{\prime})\big] \\ 
\mathcal{F}\big[v_{LR }^{20}(p,p^{\prime})\big] - \mathcal{F}\big[ v_{LR
}^{20}(0,p^{\prime})\big] & \mathcal{F}\big[ v_{LR }^{22} (p,p^{\prime})%
\big] - \mathcal{F}\big[ v_{LR }^{22} (0,p^{\prime})\big]%
\end{array}
\right) \mathbf{t}(p^{\prime },p_{0};E)  \notag \\
&&  \label{eq:9}
\end{eqnarray}

\noindent The second row of the kernel of this equation simplifies
considerably since 
\begin{equation}
\mathcal{F}[v_{LR}^{20}(0,p_0)]=\frac{v_{LR}^{20}(p_0,p_0)}{p_0^2},
\end{equation}
due to the threshold behavior of $v_{LR}$. A similar result applies to $%
v_{LR}^{22}$. Thus we also have 
\begin{equation}
\mathcal{F}[t_{20}(0,p_0;E)]=\frac{t_{20}(p_0,p_0;E)}{p_0^2}.
\end{equation}
Now, letting $E\rightarrow 0$ (i.e. $p_{0}\rightarrow 0$) in Eq.~(\ref{eq:9}%
), and extracting the $S \rightarrow D$ element of the resulting equation,
we can also employ (again because of threshold behavior) 
\begin{equation}
 \lim_{p_{0}\rightarrow 0}\mathcal{F}[v^{20}_{LR}(p,p^{\prime})]=\frac{v^{20}_{LR}(p,p^{\prime})}{%
p^2},
\end{equation}
together with the analogous result for $v^{22}_{LR}$, in order to obtain 
\begin{eqnarray}
\frac{t_{20}(p,0;0)}{p^{2}} &= & \lim_{p_{0}\rightarrow 0} \left[\frac{
t_{20}(p_{0},p_{0};E)}{p_{0}^{2}}\right]+\frac{v_{LR }^{20}(p,0)}{p^{2}}
-\lim_{p_{0}\rightarrow 0}\left[ \frac{v_{LR }^{20}(p_{0},p_{0})}{p_{0}^{2}}%
\right] \\
-\frac{2}{\pi }M\int_{0}^{\Lambda } & dp^{\prime }& \Bigg[ \left( \frac{%
v_{LR }^{20}(p,p^{\prime })}{p^{2}}-\lim_{p_{0}\rightarrow 0} \left[\frac{
v_{LR }^{20}(p_{0},p^{\prime })}{p_{0}^{2}}\right] \right) t_{00}(p^{\prime
},0;0)  \notag \\
&+& \left( \frac{v_{LR }^{22}(p,p^{\prime })}{p^{2}}-\lim_{p_{0}\rightarrow
0} \left[\frac{v_{LR }^{22}(p_{0},p^{\prime })}{p_{0}^{2}}\right] \right)
t_{20}(p^{\prime },0;0) \Bigg].  \label{eq:10}
\end{eqnarray}

This is the key result of our derivation, since it provides a second
integral equation that, together with Eq.~(\ref{eq:7.6}), is a closed set
for the half-off-shell t-matrix at zero energy. These equations require as
input the long-range potential $v_{LR}$, a choice of $C_2$, and the
experimental observables $a_t$ and the value of 
\begin{equation}
\lim_{p_{0}\rightarrow 0}\left[\frac{t_{20}(p_{0},p_{0};E)}{p_{0}^{2}}\right]%
.  \label{eq:genscat}
\end{equation}
This limit, and all the limits of $v_{LR}$ in Eq.~(\ref{eq:10}), are well
defined due to the threshold behavior of the potential and the t-matrix. The
limit (\ref{eq:genscat}) is a specific case of a ``generalized scattering
length", which is defined as 
\begin{eqnarray*}  \label{eq:105}
\frac{\alpha _{l^{\prime }l}}{M}=\lim_{p_{0}\rightarrow 0} \left[\frac{%
t_{l^{\prime }l}(p_{0},p_{0};E)}{p_{0}^{l^{\prime }+l}}\right].
\end{eqnarray*}
Since $\alpha_{l^{\prime}l}$ is related to an on-shell element of the
t-matrix it can be obtained from an experimental phase-shift analysis (see,
e.g. Ref.~\cite{PVRA05B}).

Equations~(\ref{eq:7.6}) and (\ref{eq:10}), together with an initial guess
for the value of $C_2$, now yield  $t_{00}(p,0;0)$ and $t_{20}(p,0;0)$ by
standard methods for the solution of integral equations. The other elements
of the zero-energy $\mathbf{t}$, $t_{02}(p,0;0)$ and $t_{22}(p,0;0)$, are
equal to zero, as mentioned above.

It is formally possible to eliminate $C_2$ from Eq.~(\ref{eq:9}) by a
similar set of manipulations to the ones that led to Eq.~(\ref{eq:10}). This
yields a pair of coupled integral equations for $t_{00}(p,0;0)$ and $%
t_{20}(p,0;0)$ that is NN LEC free. However, these equations contain the
quantity $\lim_{p_{0}\rightarrow 0}\mathcal{F}\big[t_{00}(0,p_{0};E)\big]$ 
as part of the driving term. This is a limit of a half-off-shell matrix
element, and so cannot be obtained in a model-independent way from
experimental data.

Therefore we now proceed similarly to the case of the momentum-dependent
interaction in the singlet channel. We guess a value for $C_2$, and solve
Eqs.~(\ref{eq:7.6}) and (\ref{eq:10}) for the half-off-shell $\mathbf{t}$ at
zero energy. We then insert these quantities into the consistency condition
for the LEC $\lambda_t$:  
\begin{equation}
\lambda _{t}=\frac{t_{20}(p,0;0)-v_{LR }^{20}(p,0)+\frac{2}{\pi }
M\;\int_{0}^{\Lambda }dp^{\prime }\;\left( v_{LR }^{20}(p,p^{\prime
})t_{00}(p^{\prime },0;0)+v_{LR }^{22}(p,p^{\prime })t_{20}(p^{\prime
},0;0)\right) }{p^{2}(1-\frac{2}{\pi }M\;\int_{0}^{\Lambda }dp^{\prime
}t_{00}(p^{\prime },0;0))}.  \label{eq:lambdat}
\end{equation}
Once we have $\lambda_t$, the consistency condition for $\lambda$ reads 
\begin{equation}
\lambda =\frac{a_{t}/M-v_{LR }^{00}(0,0)+\frac{2}{\pi }M\;\int_{0}^{\Lambda
}dp^{\prime }\;[\left( C_2 p^{\prime 2}+v_{LR }^{00}(0,p^{\prime
}))t_{00}(p^{\prime },0;0)+(\lambda _{t}p^{\prime 2}+v_{LR
}^{02}(0,p^{\prime }))t_{20}(p^{\prime },0;0)\right) ]}{1-\frac{2}{\pi }%
M\;\int_{0}^{\Lambda }dp^{\prime }t_{00}(p^{\prime },0;0)}.
\label{eq:lambda}
\end{equation}
The derivation of these two constraints for the LECs is given in Appendix~%
\ref{appendixb}. Eqs.~(\ref{eq:lambdat}) and (\ref{eq:lambda}) define the
full potential for the coupled-channels problem for a given trial value of $%
C_2$. This potential is then inserted into the LSE and phase shifts
computed. $C_2$ is adjusted until the desired experimental datum is
reproduced with sufficient accuracy.

This method has the advantage that Eqs.~(\ref{eq:lambda}) and (\ref%
{eq:lambdat}) define functional relationships: $\lambda=\lambda(C_2;a_t,%
\alpha_{20})$, and $\lambda_t=\lambda_t(C_2;a_t,\alpha_{20})$. (The
generalized scattering length $\alpha_{20}$ enters implicitly through its
effect on the half-off-shell t-matrices appearing in the formulae for $%
\lambda$ and $\lambda_t$.) Thus, we need to perform only one-parameter fits
in order to obtain all three of the LECs that are pertinent in the S-wave
J=1 channel.

\subsection{Energy-dependent central part}

Here we could follow the method of Section~\ref{sec-1S0endep} and eliminate
all three constants from the integral equation: $\lambda$, the coefficient
of the tensor short-distance interaction, $\lambda_t$, and the coefficient
of the energy-dependent part, $\gamma$. However, it is simpler to mimic the
steps of the previous Subsection, and reconstruct the underlying contact
interaction from knowledge of the phase shifts. The manner in which this is
done is described in detail in Appendix~\ref{appendixb} but can be
summarized in the following steps:

\begin{enumerate}
\item Use knowledge of $a_t$, the triplet scattering length, to eliminate $%
\lambda$ from the integral equation. Hence obtain $t_{00}(p,0;0)$.

\item Substitute the result into Eq.~(\ref{eq:7.5}), with $\gamma=0$ (since
the energy-dependent piece of the contact interaction has no impact on $t$
at $E=0$) and rearrange the equation so as to obtain $\lambda_t$ from the
on-shell quantity $\lim_{p_0 \rightarrow 0} t_{20}(p_0,p_0;E)/p_0^2 \equiv
\alpha_{20}/M$.

\item The value of $\lambda$ can then be obtained from the requirement that $%
t_{00}(0,0;0)=a_t/M$.

\item Use the ${}^3$S$_1$ and ${}^3$D$_1$ phase shifts and the mixing
parameter $\epsilon_1$ at a specific value of the energy, say $E^*$, to
reconstruct $t(p_0^*,p_0^*;E^*)$. Then, from a suitably modified (\ref{eq:7}%
), solve for $t(p,p_0^*;E)$.

\item With this half-off-shell $t$ at a finite energy in hand, $\gamma$ can
be obtained from the version of Eq.~(\ref{eq:6}) that is suitable for the
energy-dependent contact interaction. The final formula for $\gamma$ is
given in Eq.~(\ref{eq:22}).
\end{enumerate}

It may appear that in this procedure one needs five pieces of input data
to fix the constants ($a_t$, the generalized scattering length $\alpha_{20}$%
, and the three phase shifts at energy $E^*$). However, only one of the four
matrix elements of $t$ is used to extract $\gamma$. Hence, in reality only
three experimental inputs: $t_{00}(0,0;0)$, $t_{00}(p_0^*,p_0^*;E^*)$, and
the derivative of $t_{20}(p_0,p_0;E)$ with respect to $E$ at a specific energy, are needed to
solve this problem.

We then obtain phase shifts for arbitrary $E$ by by substituting the
obtained values of $\lambda $, $\lambda _{t}$ and $\gamma $ into the
potential that appears in the LSE and solving that equation for the on-shell 
$t$-matrix.

\section{Results: the Singlet S-wave}

\label{sec-result1}

In this section we present the results obtained in the ${}^1$S$_0$ channel
when we employ the three different types of contact terms introduced in
Sect.~II. First, we only adopt the leading-order contact term, i.e. a
constant. Then, we consider the energy-dependent contact term, i.e., $%
v_{SR,0}=\lambda +\gamma E$. Finally, we use the more standard
momentum-dependent contact term $v_{SR,0}=\lambda +C_2 (p^{2}+p^{\prime 2})$%
. For each of these cases, we examine the results found with the following
three different forms of the long-range potential $v_{LR}$:

1. TPE computed using dimensional regularization up to $O(P^2)$ (denoted as
DR NLO).

2. TPE computed using dimensional regularization up to $O(P^3)$ (denoted as
DR NNLO).

3. TPE computed using spectral-function regularization up to $O(P^3)$
(denoted as SFR NNLO).

For the case of spectral-function regularization, we adopt $\bar{%
\Lambda}=800$~MeV as intrinsic cutoff. The effect that varying the intrinsic cutoff from $600-800$ MeV has on this potential is discussed in reference \cite{PVRA09}. When combined with the
momentum-dependent contact interaction, the $O(P^2)$ TPE yields the usual
NLO calculation in $\chi$ET, while the two $O(P^3)$ forms of $v_{LR}$ both
give calculations of NNLO accuracy.

\subsection{The lowest-order contact term}

A proper renormalization of these TPE potentials entails the presence of
contact terms up to second order in $|\mathbf{q}|$. However, it is still of
interest to investigate what happens if only the lowest-order contact term
is included in the potential (see also Refs.~\cite{PVRA06A,En08}). This
enables us to see how the phase shift converges as the LSE cutoff $\Lambda$
is increased. Then, when higher-derivative contact interactions are added to
the potential, we can see how the cutoff dependence changes.

The renormalization for the leading-order contact term is performed with one
subtraction, with the scattering length $a_{s}=-23.7$ fm~\cite{nnonline}
used as renormalization condition. This is the same as procedure as
described for the ${}^1$S$_0$ channel in Ref.~\cite{Ya08}, except that here
we are using a long-range potential that is higher-order in the chiral power
counting.

\begin{figure}[tbp]
\begin{center}
\includegraphics[width=8cm]{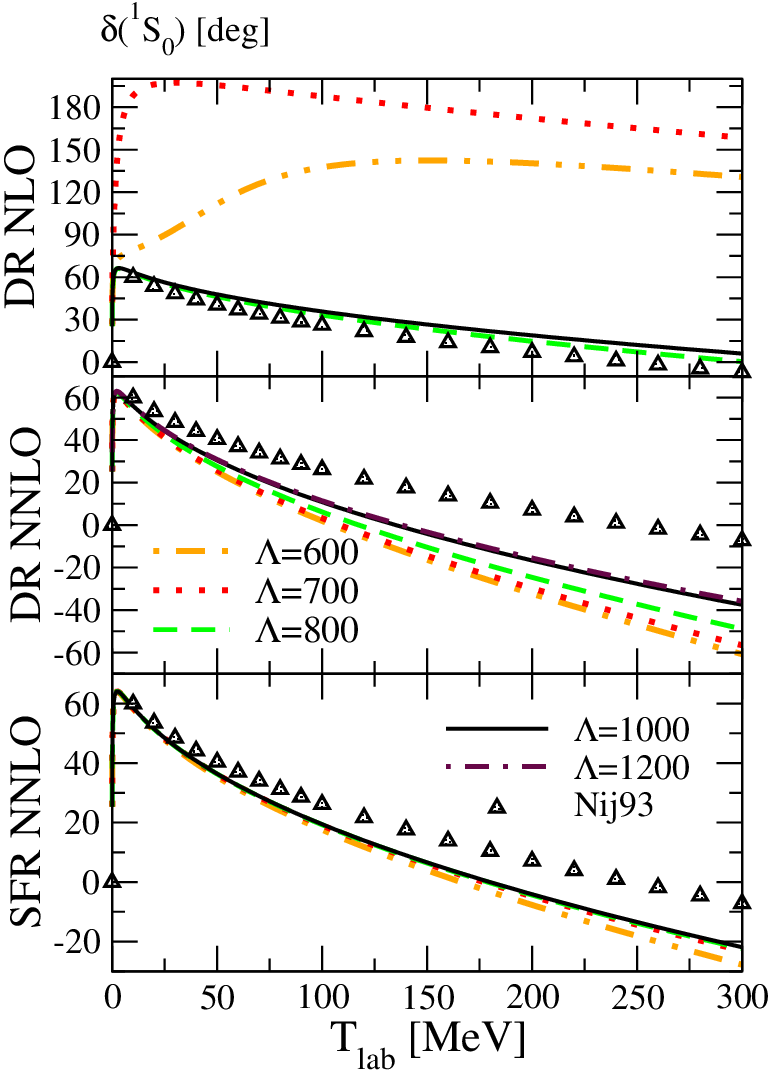}
\end{center}
\par
\vspace{3mm}
\caption{(Color online) The $^1$S$_0$ NN phase shift as a function of the
laboratory kinetic energy for different cutoffs $\Lambda$ ranging from 0.6
to 1.2~GeV. The long-range potentials employed are DR NLO, DR NNLO and SFR
NNLO (as noted in the y-axis of each figure), together with a constant
contact term. The results are obtained by one subtraction with $a_0=-23.7$
fm as input. The values of the Nijmegen phase-shifts~\protect\cite{nnonline}
are indicated by the open triangles.}
\label{fig-fig1}
\end{figure}

The phase shifts $^1$S$_0$ calculated with the DR NLO, DR NNLO and SFR NNLO
potential are shown for different cutoffs $\Lambda $ in Fig.~\ref{fig-fig1}.
The results show that a single, constant, contact interaction is sufficient
to stabilize the $^1$S$_0$ phase shift with respect to the cutoff once $%
\Lambda >800$~MeV for the SFR NNLO TPE and $\Lambda >1000$ MeV for the DR
NLO TPE. The DR NNLO potential requires $\Lambda >1200$ MeV to become stable.

The DR NLO potential exhibits a resonance-like behavior when $\Lambda \leq
700$ MeV, but the phase shift quickly converges after $\Lambda $ exceeds
800~MeV. This indicates that---at least for this potential---a cutoff
smaller than $700$~MeV removes essential physics, and brings the phase shift
rather close to that obtained with $v_{LR}=v_{OPE}$ in Ref.~\cite{Ya08}. On
the other hand, it is amusing to note that after $\delta(^1$S$_0)$ converges
with respect to the cutoff, the DR NLO result is actually closer to the
PWA93 phase shifts~\cite{St93,nnonline} than is the result from the other
two potentials.

This indicates that the pure (i.e. un-compensated by contact interaction)
NNLO TPE produces too large a correction, although the general trend of
providing more repulsion is correct. We note that the phase shift calculated
with the SFR NNLO potential stabilizes as a lower value of $\Lambda$, and
that this $\Lambda \rightarrow \infty$ result is closer to the PWA93 result,
compared to that obtained with the DR NNLO potential.

\begin{figure}[tbp]
\begin{center}
\includegraphics[width=16cm]{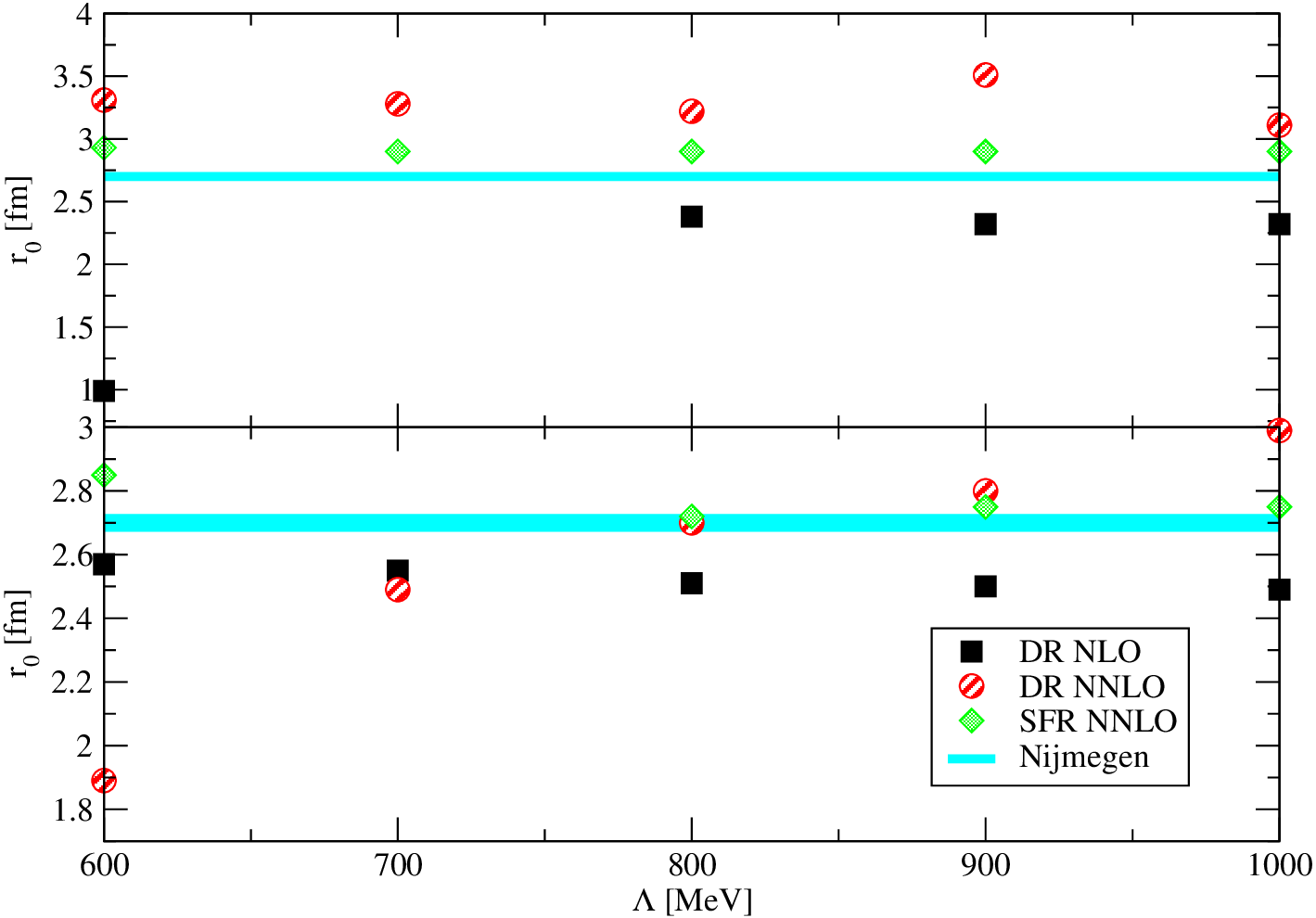}
\end{center}
\par
\vspace{3mm}
\caption{(Color online) The effective range $r_0$ [in fm] in the $^1$S$_0$ channel
extracted from calculations with the DR~NLO (black square), the DR~NNLO (red circle),
 and the SFR~NNLO (green diamond) TPE combined with: a constant contact term (upper panel), and a constant plus a momentum-dependent contact term (lower panel). 
In both cases $r_0$ is shown as a function of the
cutoff $\Lambda$ in the LSE. In the lower panel, the coefficient of the momentum-dependent contact term is adjusted to reproduce the Nijmegen value of the phase shift at $T_{lab}=200$ MeV. The thick solid band represents the range of $r_0$ obtained from Ref.\cite{PVRA05B}. Note that for the upper(lower) panel, the value of $r_0$ at $\Lambda=700$ MeV for the DR NLO (SFR NNLO) potential is $-10.2(0.15)$ fm, which is not plotted in the figure. }
\label{fig-t1}
\end{figure}

In the upper panel of Fig.~\ref{fig-t1} we plot the values of the effective range $r_{0}$
obtained from the above potentials. Since $a_s$ is fixed to the same value
in all calculations, this figure shows the influence of $\Lambda$ on the
low-energy phase shifts. Comparing our results for the three different
potentials to the ``experimental'' value $r_{0}=2.7$~fm~\cite{nnonline}, we
see that the SFR~NNLO potential gives a value of $r_{0}$ that is closest to
the experimental value. The results with the DR NLO and SFR NNLO TPE are
stable with respect to $\Lambda$ once $\Lambda \geq 700$~MeV. The DR NNLO
TPE value eventually also stabilizes within small variations once $\Lambda >
1200$~MeV (not shown in Fig.~\ref{fig-t1}). This cutoff is rather
large, reflecting the too-strong energy dependence that was already
discussed in the previous paragraph.

\subsection{Second order: energy-dependent contact term}

\label{sec-results1S0endep}

Next we use a second-order contact term with the simplest energy-dependent
form, namely $v_{SR,0}=\lambda +\gamma E$. This form arises because the
contact operator $N^\dagger i \partial_t N N^\dagger N$ can absorb the
divergences in the two-pion exchange potential, with the difference absorbed
into terms that are proportional to the free-nucleon equation of motion.
(See, e.g. Ref.~\cite{Ka97}, where the amplitude is calculated in Born
approximation, and so the contact terms required are only a function of the
nucleon on-shell momentum $p_0$.) These additional terms can be transformed
into higher-order multi-body operators. Thus, up to $O(P^3)$ in $\chi$PT,
the energy-dependent contact term is just as valid a representation of the
short-distance physics as the more commonly chosen momentum-dependent one.
Of course, the use of energy-dependent potentials engenders additional
complications---especially when one wishes to embed the two-nucleon Hilbert
space in a larger Fock space. Nevertheless, we can choose this form for the
short-distance operators to describe NN scattering, and it has the advantage
that the two unknown constants can be solved by our double-subtraction
method discussed in Sect.~II.

As input this method requires the value of $a_{s}$ and the phase shift at a
specific energy $E^{\ast }$. In principle, this energy can be chosen
arbitrarily. In our calculation we choose $T_{lab}=2E^{\ast }=2.8$ MeV, i.e.
at the energy where the $^{1}S_{0}$ phase shift has its turning point. We
tested other choices $E^{\ast }<10$~MeV and found that they have only a minor
effect on the results. We also tested a relatively high energy, $T_{lab}>100$%
~MeV. Such a choice generates results which fit the phase shift at low
energies (due to $a_{s}$ as input) and at the chosen higher energy (due to
the use of $\delta(E^{\ast })$ as input). However, we did not adopt this
procedure, since in it the ultimate dependence of results on the choice of
renormalization point (i.e. $E^*$) reflects the the cutoff-dependence in the
phase shift in a manner that we shall discuss below.

\begin{figure}[tbp]
\begin{center}
\includegraphics[width=16cm]{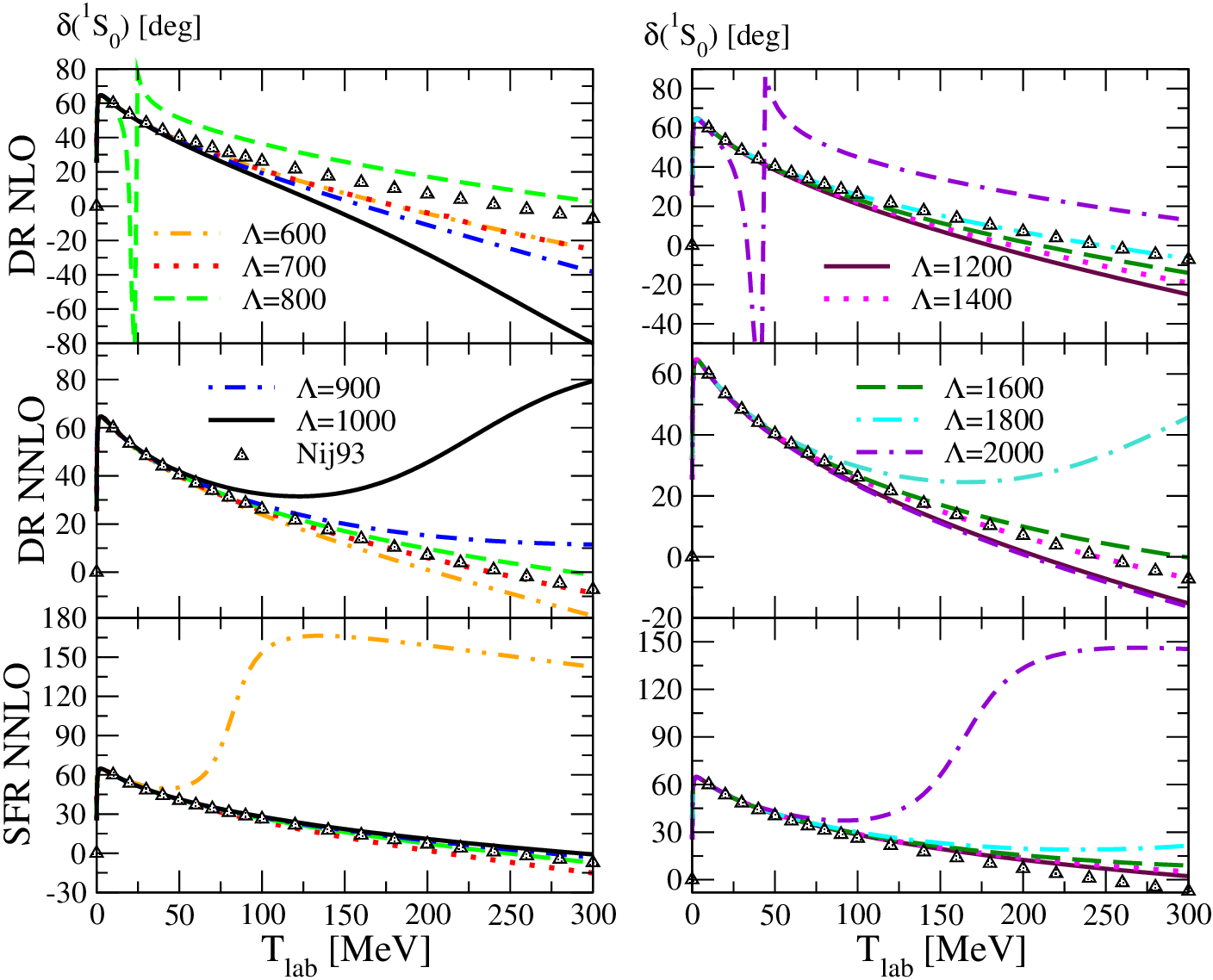}
\end{center}
\par
\vspace{3mm}
\caption{(Color online) The $^1$S$_0$ NN phase shift as a function of the
laboratory kinetic energy for different cutoffs $\Lambda$ ranging from 0.6
to 2~GeV. The potentials employed (from top to bottom) are DR NLO, DR NNLO
and SFR NNLO, together with an energy-dependent contact term. The results
are obtained by two subtractions with $a_0=-23.7$ fm and the phase shift at $%
T_{lab}=2.8$ MeV as input. The values of the Nijmegen phase-shifts~%
\protect\cite{nnonline} are indicated by the open triangles.}
\label{fig-fig44}
\end{figure}

In Fig.~\ref{fig-fig44} the $^1$S$_0$ phase shift calculated with the TPE
potentials DR~NLO, DR~NNLO and SFR~NNLO with the energy-dependent contact
term is displayed for cutoffs $\Lambda =600-2000$~MeV. A general feature is
that the phase shift oscillates as a function of $\Lambda$. There are also
strongly diverging phase shifts at $\Lambda =800$ and $2000$ MeV for the
DR~NLO potential because a resonance state is created by the contact term.
(A similar behavior for the DR~NLO potential was observed in Ref.~\cite{En08}
for a momentum-dependent contact term with $\Lambda =2000$~MeV.) For the DR
NNLO (SFR NNLO) TPE, at $\Lambda=1000 (2000)$~MeV the phase shift diverges
violently from the previous converged value.

\begin{figure}[tbp]
\begin{center}
\includegraphics[width=16cm]{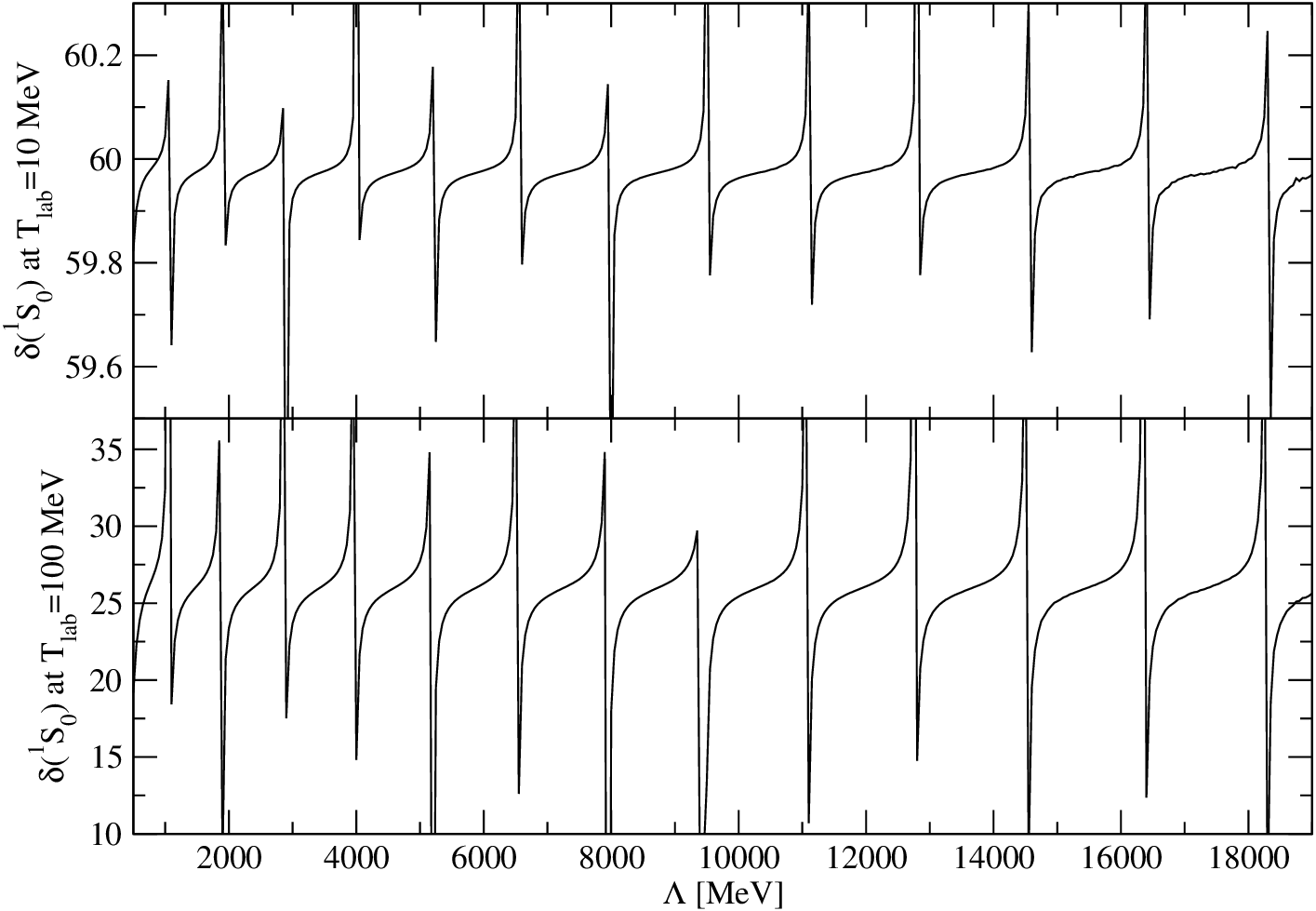}
\end{center}
\par
\vspace{3mm}
\caption{The $^1$S$_0$ NN phase shift at $T_{lab}=10$ MeV (upper panel) and $100$ MeV (lower panel) as a function of
the cutoff ranging from 0.5--19~GeV. The results are obtained using the DR
NNLO potential with an energy-dependent contact term via two subtractions.}
\label{fig-fig7}
\end{figure}

In order to further understand the effect of the energy-dependent contact
term, we calculate the DR NNLO TPE phase shift at $T_{lab}=10$ and $100$ MeV for $%
\Lambda $ up to $19$ GeV and display the results in Fig.~\ref{fig-fig7}. The
phase shifts have the same oscillatory behavior, with the period becoming longer as $%
\Lambda $ becomes larger. Comparison with the results obtained by adopting
only a constant contact term (second panel of Fig.~\ref{fig-fig1}, where
this phase shift converges with respect to $\Lambda$ for $\Lambda >1200$
MeV), suggests that the quasi-periodic behavior of $\delta(T_{lab}=10$ \& $100~%
\mathrm{MeV})$ is caused by the energy-dependent contact term.

The oscillatory behavior of the phase shift at this energy is associated
with resonances at specific $\Lambda$. An argand plot confirms that there is
a resonance for the cutoff at which this phase shift diverges. Tracking the
energy of the resonance as a function of $\Lambda$ shows that it moves in a
regular way, with the real and imaginary parts of $E_R$ decreasing in size
as $\Lambda$ increases, until eventually the resonance becomes a bound state
and the pole in t moves towards an energy of $-\infty$. In order to
understand this better, we perform a two-potential analysis of the t-matrix
resulting from this particular $v_{SR}$. The details are given in Appendix~%
\ref{appendixc}. We find that the appearance of poles is a consequence of
the specific form of the contact interaction. When these poles are in the
domain of validity of the theory the resulting phase shifts are in poor
agreement with the Nijmegen PSA. Only at those cutoffs where the artificial
poles correspond to deeply bound states or high-energy resonances do the
phase shifts match the Nijmegen analysis well.

The effective range $r_{0}$ obtained from all three different potentials is
about $r_{0}=2.68$fm and varies by $< 2\%$ over the range of cutoffs
considered, with the variation being only $\sim 0.5\%$ if the long-range
part is computed to $O(P^3)$. This is a consequence of our choice of a
low-energy phase shift as second input. Comparing with the upper panel of Fig.~\ref{fig-t1},
one can clearly see that $r_{0}$ receives significant correction from the $%
O(P^{2})$ contact term, especially for the DR~NLO potential.

Thus, the energy-dependent contact term improves the overall agreement of
the ${}^1$S$_0$ phase shift, and in particular the effective range $r_{0}$,
with respect to the average value $r_{0}=2.7$~fm extracted from the
different Nijmegen potentials. Once one tries to increase the cutoff $%
\Lambda $ beyond $1000$ MeV, this form of short-distance physics leads to
problems if the DR TPE is used as long-range potential. However, adopting
TPE computed using spectral-function regularization means a wider range of
cutoffs can be employed before the same issues occur as for the DR
TPE.

It is worthwhile mentioning that there exists a particular regularization of
two-pion exchange which avoids resonant behavior in the phase shifts. This
involves using dimensional regularization for the NLO part of TPE, and
spectral-function regularization for the NNLO part\footnote{see \cite{Ya09} for the detail of this potential and the analysis in p-waves.}. Then, even if an
energy-dependent contact term is chosen for the short-range potential, only
minor oscillations of the phase shifts with respect to $\Lambda$ occur for $%
\Lambda > 1$ GeV. However, strictly speaking, a potential which is obtained
using two different regularization schemes is not self-consistent. Thus, in
the later discussion we will only present the results obtained from the
three TPE potentials mentioned at the beginning of this section.

\subsection{Momentum-dependent contact term}

A momentum-dependent contact term associated with the TPE has the form $%
\lambda +C_2 (p^{2}+p^{\prime 2})$, and can be solved by one subtraction
plus the use of a one-parameter fit, as explained in Sect.~II. The input
chosen for the subtraction is again $a_s$. In what follows we will try to
perform two different fits to fix the value of $C_2$. One uses only
low-energy information, namely the effective range, $r_{0}=2.7$~fm. The
other fit fixes $C_2$ by attempting to reproduce the phase shift at $%
T_{lab}=200$~MeV.

\begin{figure}[tbp]
\begin{center}
\includegraphics[width=9cm]{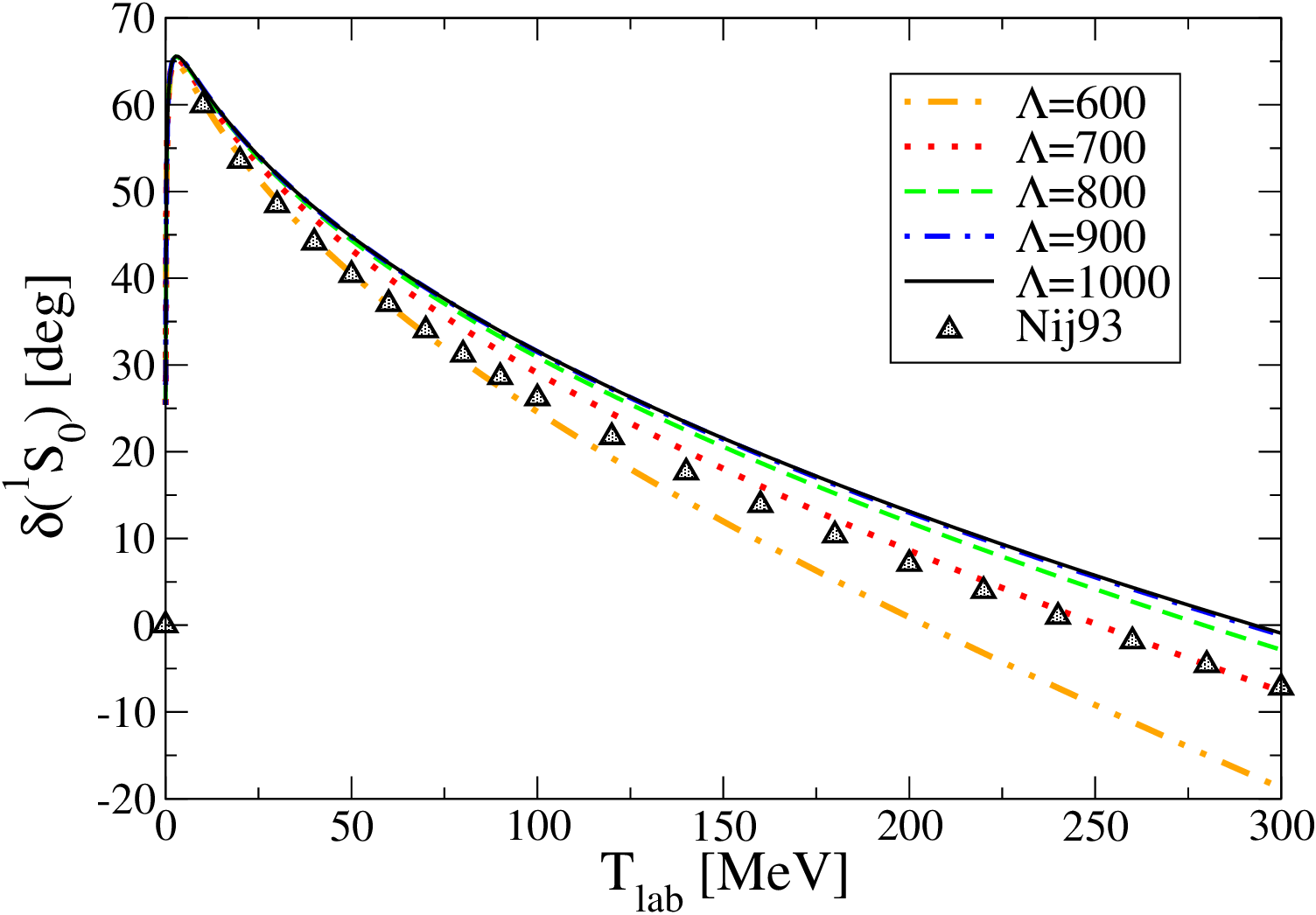}
\end{center}
\par
\vspace{3mm}
\caption{(Color online) The $^1$S$_0$ NN phase shift as a function of the
laboratory kinetic energy for different cutoffs $\Lambda$ ranging from 0.6
to 1~GeV. The DR NLO long-range potential is employed, together with a
momentum-dependent contact term. The results are obtained by making one
subtraction with $a_0=-23.7$ fm as input, and then performing a best fit to
the overall phase shift as given by the Nijmegen analysis. The values of the
Nijmegen $^1$S$_0$ phase shifts~\protect\cite{nnonline} are indicated by the
open triangles.}
\label{fig-fig8}
\end{figure}

The resulting $^1$S$_0$ phase shift is shown in Fig.~\ref{fig-fig8} for the
DR NLO TPE. In this case we cannot reproduce $r_0$, or the phase shift at $%
T_{lab}=200$ MeV, once $\Lambda > 600$ MeV. This pathology of the fit at NLO
was discussed in Ref.~\cite{PVRA06A}, and is related to the Wigner bound
that limits the impact of short-distance physics on the effective range---or
more generally on the energy dependence of phase shifts~\cite{PC96,Sc96}.
Therefore, for the DR NLO TPE we can only perform an overall best fit to the
phase shift. Fig.~\ref{fig-fig8} shows that, within this limitation, the
phase shift quickly becomes cutoff independent once $\Lambda >900$ MeV.

\begin{figure}[tbp]
\begin{center}
\includegraphics[width=16cm]{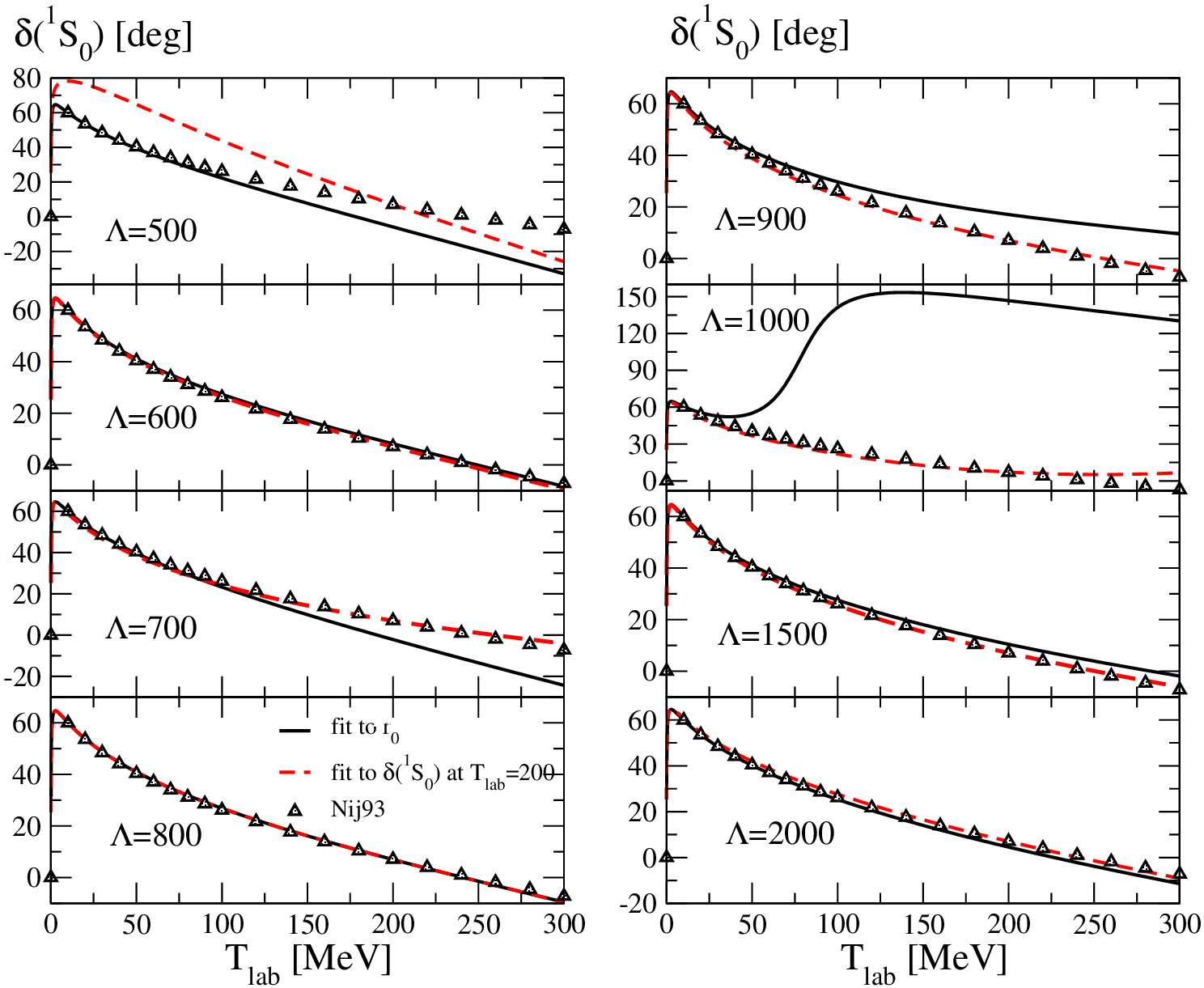}
\end{center}
\par
\vspace{3mm}
\caption{(Color online) The $^1$S$_0$ NN phase shift as a function of the
laboratory kinetic energy for different cutoffs $\Lambda$ ranging from 0.5
to 2~GeV. The potential employed is the DR NNLO TPE with a momentum-dependent
contact term. The results are obtained by making one subtraction with $%
a_0=-23.7$ fm as input and then performing a fit to either the effective
range $r_0=2.7$ fm (solid line) or the Nijmegen $^1$S$_0$ phase shift at $%
T_{lab}=200$~MeV (dashed line). The values of the Nijmegen phase-shifts~%
\protect\cite{nnonline} are indicated by the open triangles.}
\label{fig-fig9}
\end{figure}

The results for the DR NNLO TPE are shown in Fig.~\ref{fig-fig9}, where we
see that the two different fit procedures generate different results for the
same $\Lambda $. This is especially visible at $\Lambda$~=~500 and 1000~MeV,
where a resonance-like behavior is present in the latter case when $C_2$ is
fitted to $r_{0}$. For values of $\Lambda $ not close to these problematic
cutoffs the phase shift is almost independent of the renormalization point.

\begin{figure}[tbp]
\begin{center}
\includegraphics[width=16cm]{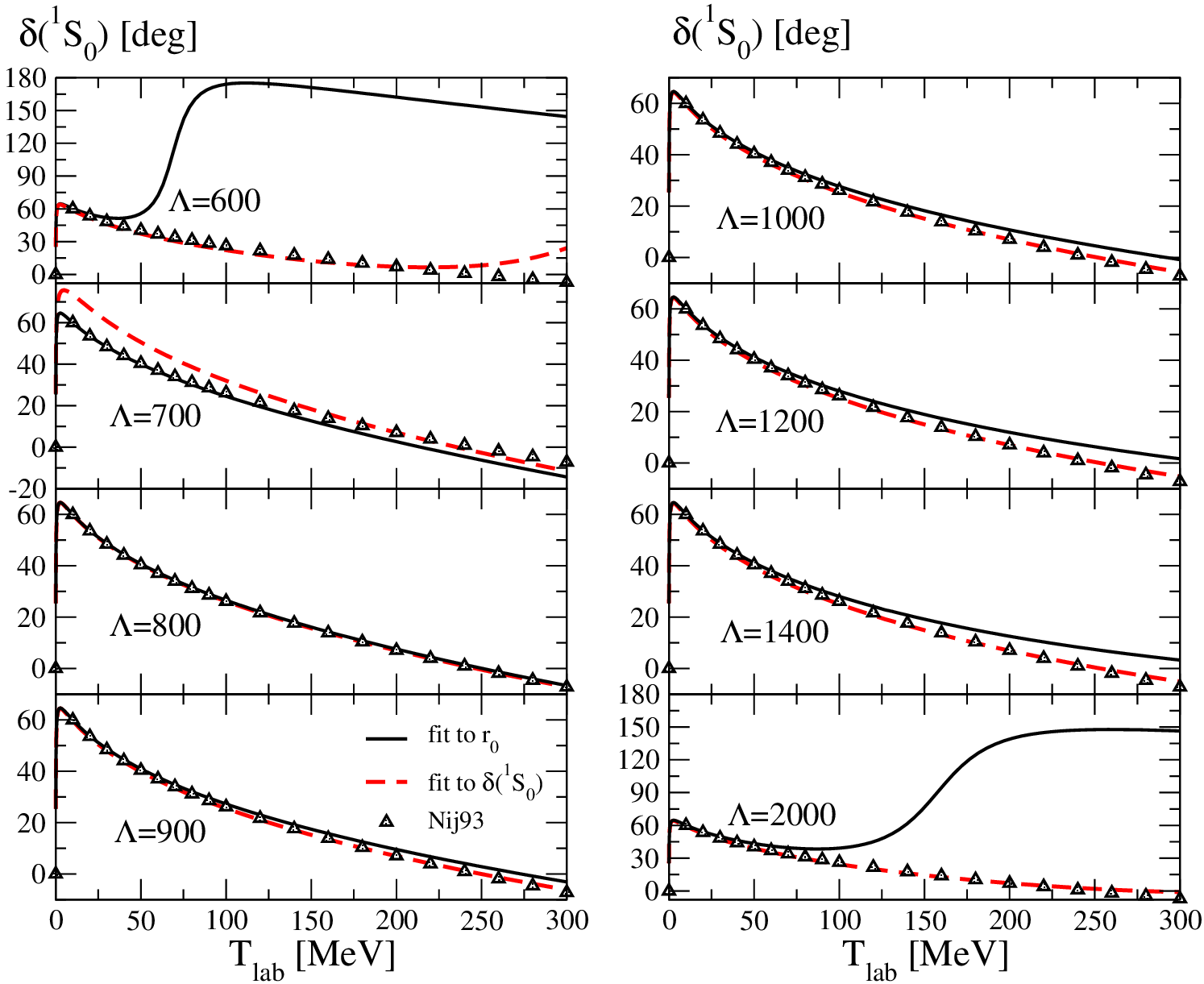}
\end{center}
\par
\vspace{3mm}
\caption{(Color online) The $^1$S$_0$ NN phase shift as a function of the
laboratory kinetic energy for different cutoffs $\Lambda$ ranging from 0.6
to 2~GeV. The potential employed is SFR NNLO with the momentum-dependent
contact term. The results are obtained by making one subtraction with $%
a_0=-23.7$ fm as input and then performing either a fit to the effective
range $r_0=2.7$ fm (solid line) or the Nijmegen $^1$S$_0$ phase shift at $%
T_{lab}=200$~MeV (dashed line). The values of the Nijmegen phase-shifts~%
\protect\cite{nnonline} are indicated by the open triangles.}
\label{fig-fig100}
\end{figure}

This independence of renormalization point is even more apparent when the
SFR NNLO TPE is adopted. As can be seen in Fig.~\ref{fig-fig100}, for $%
\Lambda$ between $700-1800$~MeV, the two different fitting procedures lead
to almost the same $^1$S$_0$ phase shift. The same figure also suggests that 
$\Lambda < 600$ MeV is too small and cuts off too much of the potential to
allow a good fit to the Nijmegen analysis. Finally, the agreement between
the two fits breaks down at $\Lambda=2000$ MeV. Thus, by adopting the SFR NNLO
TPE, we achieve renormalization-point independence for a wider range of
cutoffs, but this renormalization-point independence breaks down for higher 
$\Lambda$.

Finally, we show in the lower panel of Fig.~\ref{fig-t1} the extracted effective range $r_{0}$
as function of  $\Lambda $ for the $^1$S$_0$ phase shift fitted to the
Nijmegen value at $T_{lab}=200$ MeV. This allows to assess the sensitivity
of the results to renormalization point ($E^* \approx 0$ or $E^* \approx 200$
MeV lab. energy). The DR NLO TPE has the least variation of $r_{0}$ with $%
\Lambda$ over the range of $\Lambda$ considered here, but it needs to be
remembered that the experimental value cannot be reproduced with a real
value of $C_2$~\cite{PVRA06A,PVRA08A}. At order NNLO, the SFR TPE gives
values of $r_{0}$ which vary less with $\Lambda$ than do the ones from the
DR TPE once $\Lambda>800$ MeV, and are stable until $\Lambda=2000$ MeV. This
can already be inferred from Fig.~\ref{fig-fig100}.

In summary, the results obtained from the NNLO DR TPE with a
momentum-dependent contact term have a peculiar behavior at $\Lambda =1000$%
~MeV. Adopting the SFR potential increases the validity range of $\Lambda$
as in the energy-dependent case.

\section{Results in the J=1 triplet channel}

\label{sec-result2}

In the J=1 triplet channel we again adopt the three different potentials,
i.e., the DR TPE up to NLO, the DR TPE up to NNLO and the SFR TPE up to
NNLO. To renormalize the potential in this channel we employ three different
contact terms (cases A--C in Sec.~\ref{sec-triplet}). For case B
(energy-dependent) and C (momentum-dependent), the value of the generalized
scattering length 
\begin{equation*}
\lim_{p_{0}\rightarrow 0}\frac{t_{20}(p_{0},p_{0};E)}{p_{0}^{2}}\equiv \frac{%
\alpha _{20}}{M}
\end{equation*}%
is required. We first adopt the value given in \cite{PVRA05B} for $\alpha
_{20}$. Then we adjust $\alpha _{20}$ to find the best fit. Since $\alpha
_{20}$ is not given from experiment, we list the values extracted from
several so called \textquotedblleft high-precision" potentials in Table~\ref%
{table-5}.

\subsection{The lowest-order contact term}

\begin{figure}[tbp]
\begin{center}
\includegraphics[width=16cm]{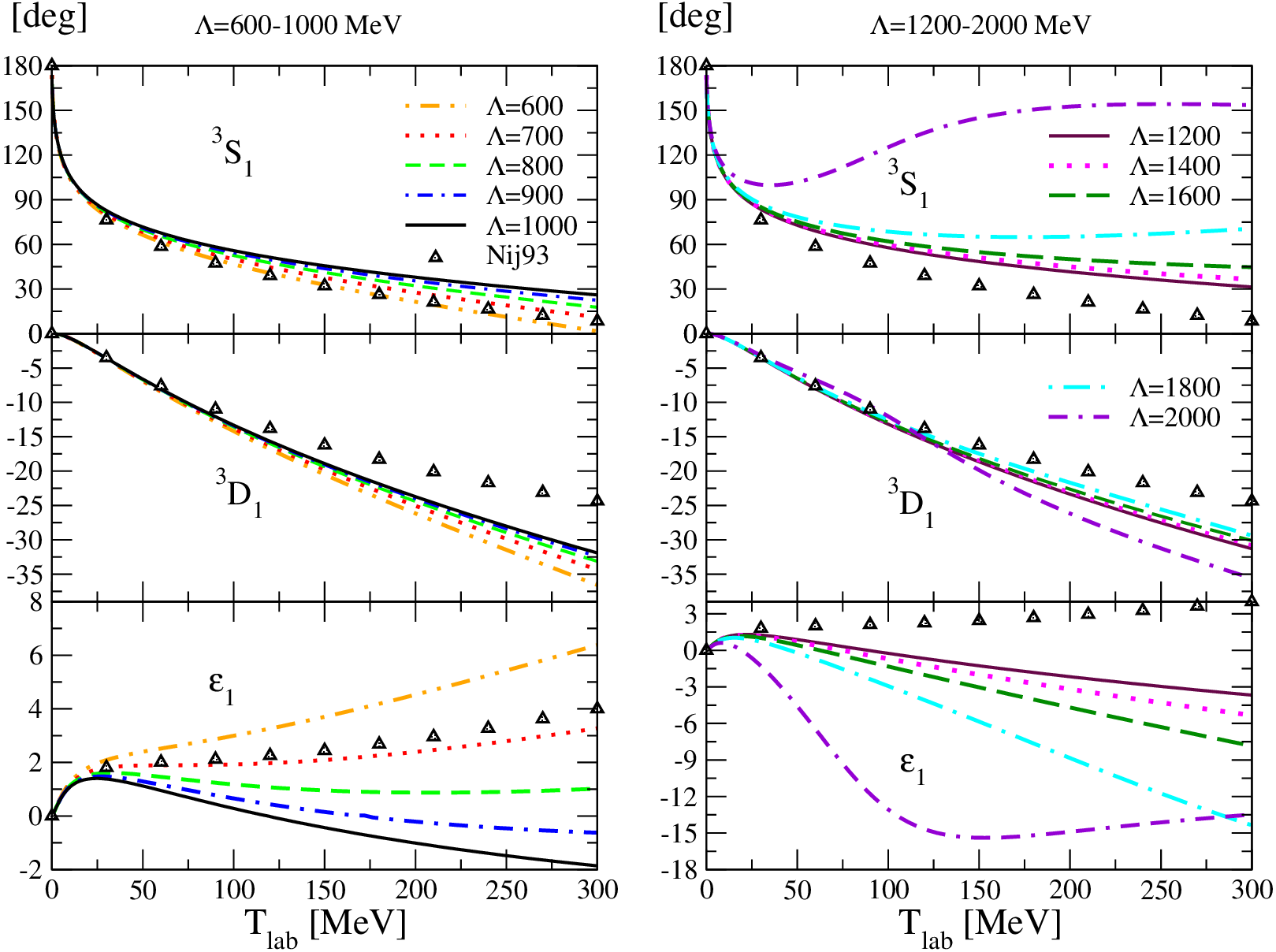}
\end{center}
\par
\vspace{3mm}
\caption{(Color online) The J=1 coupled NN phase shifts as a function of the
laboratory kinetic energy for different cutoffs $\Lambda$ ranging from 0.6
to 2~GeV. The potential employed is the DR NLO TPE with one constant contact
term. The results are obtained via one subtraction with $a_0=5.428$ fm as
input. The values of the Nijmegen phase-shifts~\protect\cite{nnonline} are
indicated by the open triangles.}
\label{fig-fig11}
\end{figure}

\begin{figure}[tbp]
\begin{center}
\includegraphics[width=16cm]{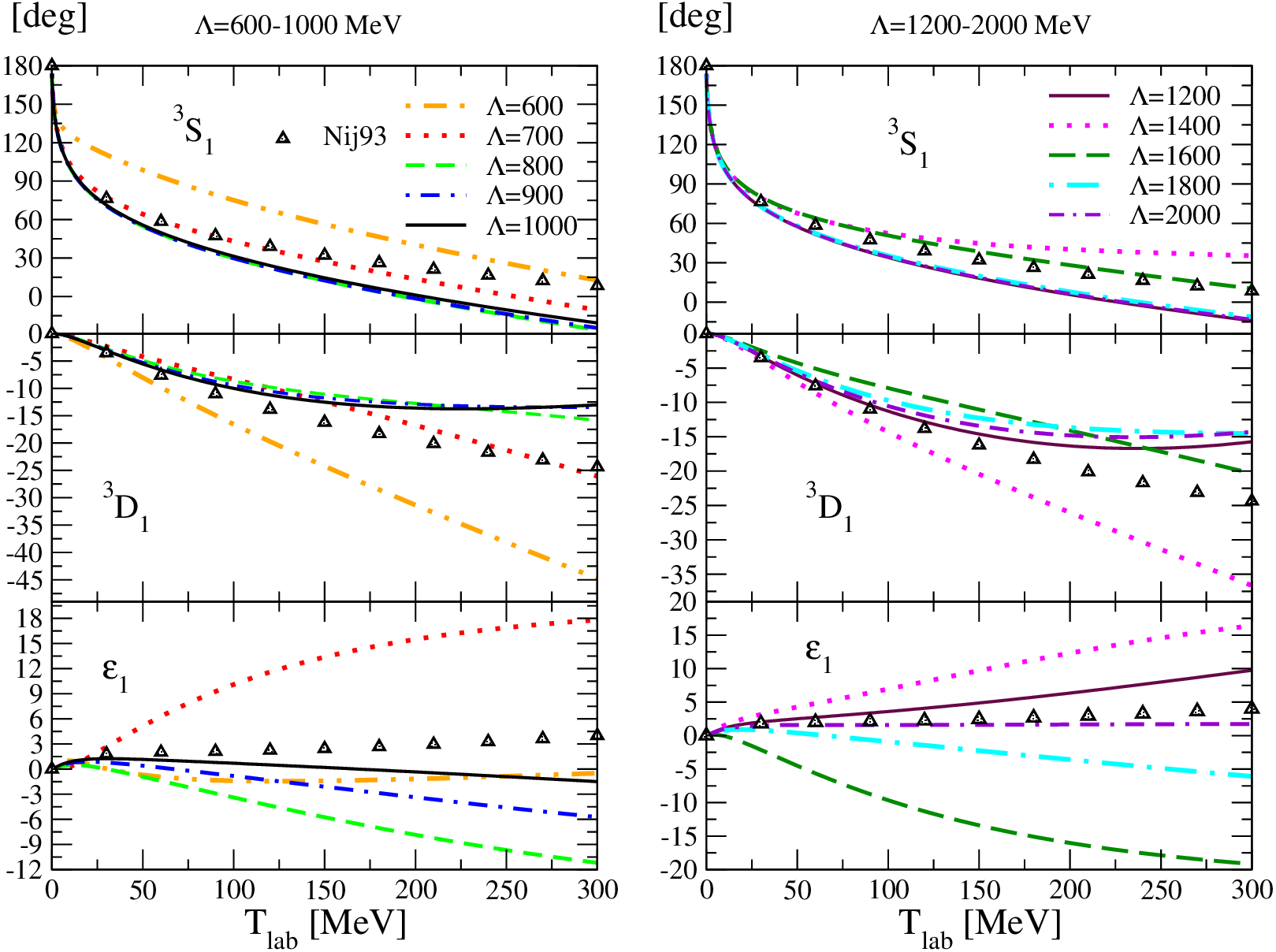}
\end{center}
\par
\vspace{3mm}
\caption{(Color online) The J=1 coupled NN phase shifts as a function of the
laboratory kinetic energy for different cutoffs $\Lambda$ ranging from 0.6
to 2~GeV. The potential employed is DR NNLO with a constant contact term.
The results are obtained via a single subtraction with $a_0=5.428$ fm as
input. The values of the Nijmegen phase-shifts~\protect\cite{nnonline} is
indicated by the open triangles.}
\label{fig-fig12}
\end{figure}

\begin{figure}[tbp]
\begin{center}
\includegraphics[width=16cm]{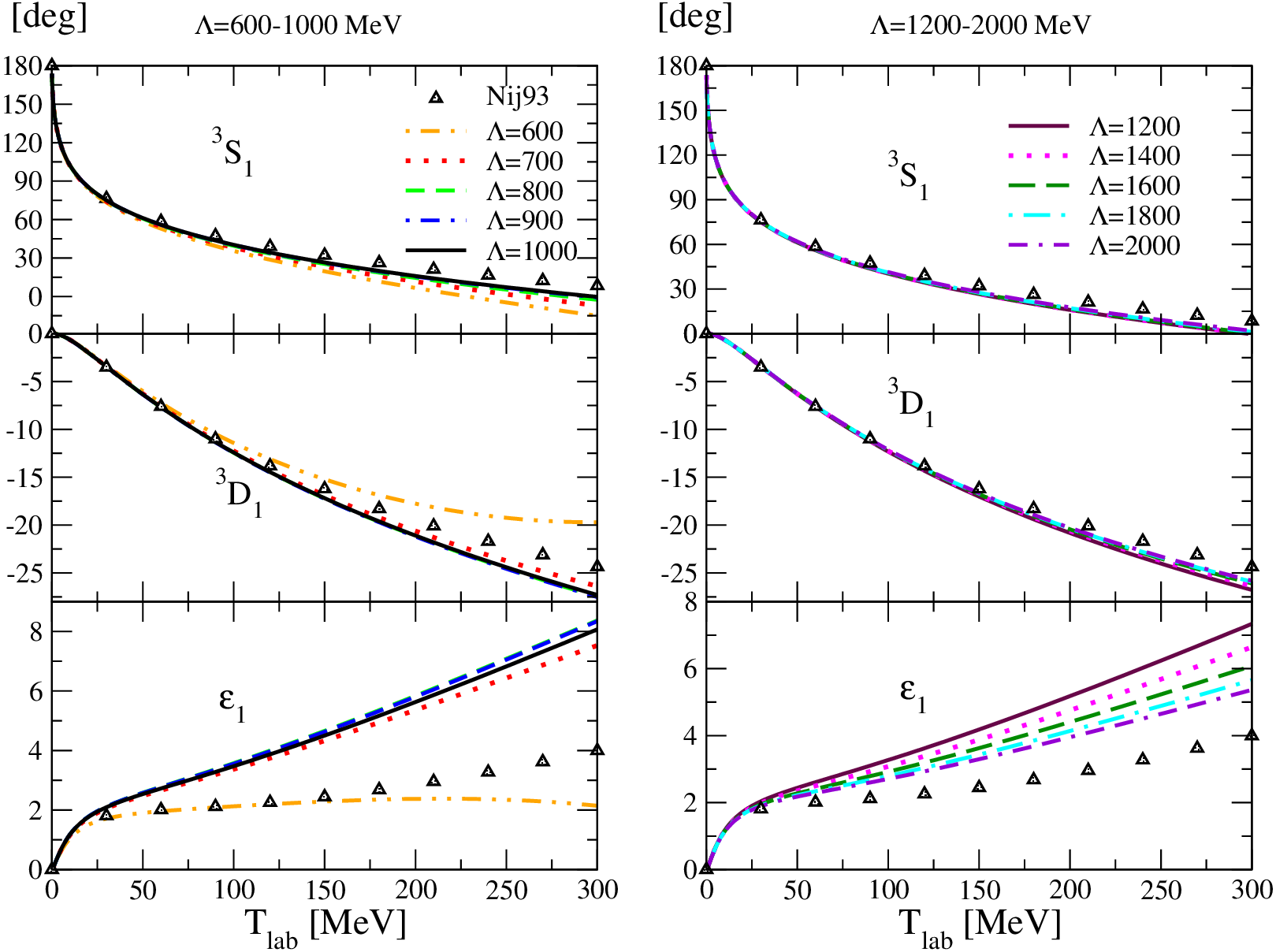}
\end{center}
\par
\vspace{3mm}
\caption{(Color online) The J=1 coupled NN phase shifts as a function of the
laboratory kinetic energy for different cutoffs $\Lambda$ ranging from 0.6
to 2~GeV. Here the potential is SFR NNLO with a constant contact term. The
results are obtained by one subtraction with $a_0=5.428$ fm as input. The
values of the Nijmegen phase-shifts~\protect\cite{nnonline} are indicated by
the open triangles.}
\label{fig-fulltc}
\end{figure}

We first perform the renormalization with the constant contact term (A) in
the $^{3}$S$_{1}-^{3}$S$_{1}$ channel. The results obtained with this
short-distance potential, and long-distance potentials DR NLO, DR NNLO and
SFR NNLO are shown in Figs.~\ref{fig-fig11}-\ref{fig-fulltc}, respectively.
In the first two cases the $^{3}$S$_{1}$ phase shifts reach cutoff
independence as $\Lambda $ approaches $1000$ MeV, and then diverge once more
at higher $\Lambda $. The $^{3}$D$_{1}$ phase shifts exhibit a similar
feature, although the pattern of convergence is not as obvious as in $^{3}$S$%
_{1}$. For the mixing angle $\epsilon _{1},$ there is no obvious convergence
with respect to $\Lambda$ for the DR NLO and DR NNLO TPE. This can be
explained by the fact that we only adopt a contact term in the $^{3}$S$%
_{1}-^{3}$S$_{1}$ channel and fit to the scattering length $a_{t}=5.428$ fm.
This was sufficient to stabilize the J=1 triplet channel in the case of the
one-pion-exchange potential\cite{Be02,PVRA04B,Ya08}. However, the NNLO TPE is singular and
attractive in both eigen-channels and thus needs two contact interactions
(or, equivalently, two subtractions) in order to stabilize its predictions
with respect to $\Lambda$~\cite{PVRA06A}. In contrast, the NLO TPE has a
repulsive behavior as $r \rightarrow 0$ in both J=1 eigen-channel. This also
leads to predictions that are unstable after one subtraction, albeit for the
opposite reason than at NNLO.

\begin{figure}[tbp]
\begin{center}
\includegraphics[width=16cm]{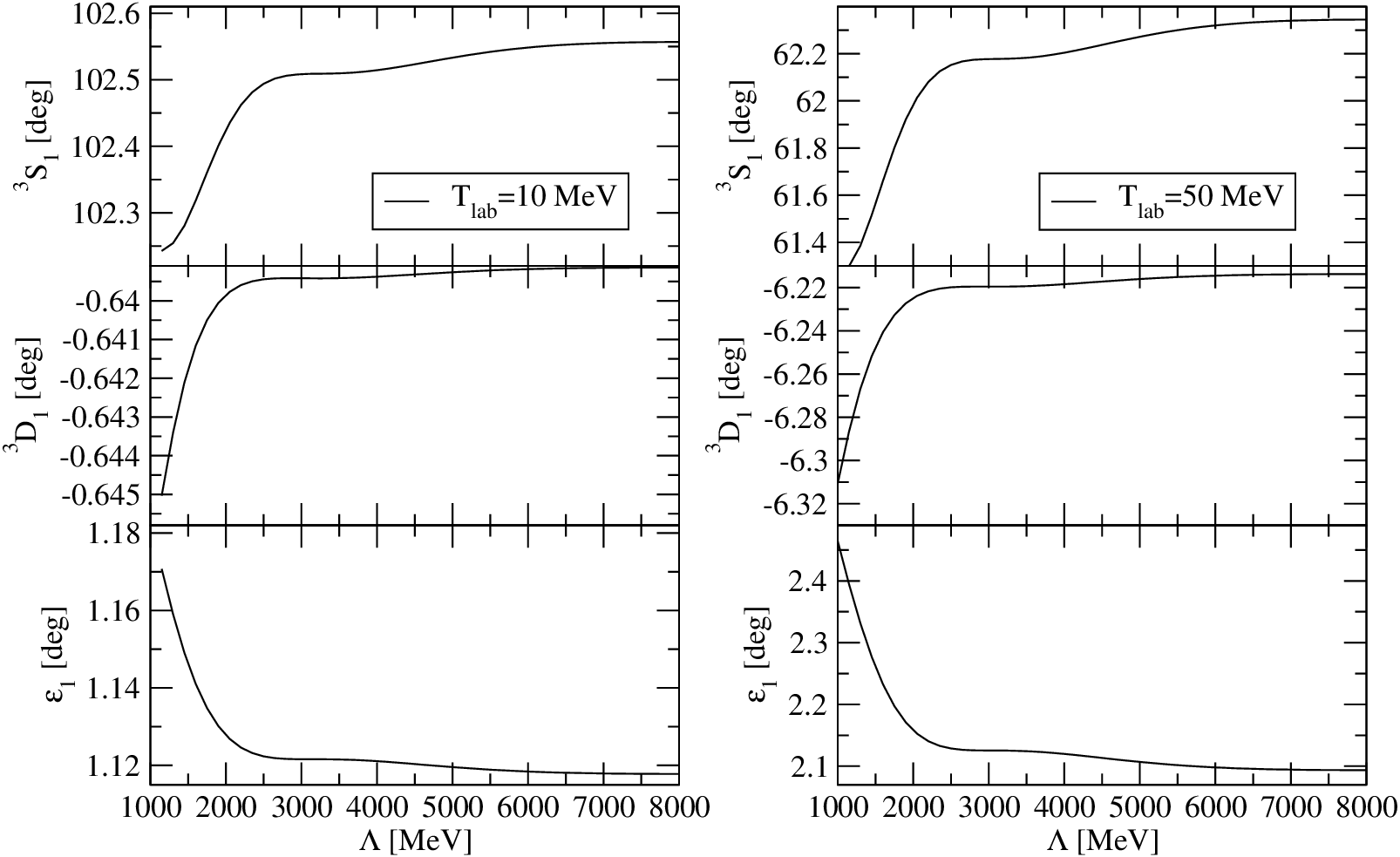}
\end{center}
\par
\vspace{3mm}
\caption{(Color online) The J=1 coupled NN phase shifts at $T_{lab}=10$ (left panel) and $50$(right panel)~MeV
as a function of cutoff ranging from 1--8~GeV. The results are obtained
with the SFR NNLO potential and a constant contact term.}
\label{fig-3sc}
\end{figure}

The SFR NNLO potentials do not have the same singular attractive behavior at
short distances, and so it is no coincidence that its phase shifts show the
best overall convergence in $^{3}S_{1}$, $^{3}D_{1}$, and $\epsilon_1$. The
SFR strongly reduces the short-range attraction in the NNLO pieces of the
TPE, and a lower-order contact term is then able to remove the $\Lambda$
dependence of the phase shifts. Fig.~\ref{fig-3sc} shows that the phase
shifts at $T_{lab}=10$ and $50$ MeV obtained with the SFR NNLO TPE and a constant
contact term become stable for $\Lambda>2500$ MeV.

\subsection{Energy-dependent contact term}

\begin{figure}[tbp]
\begin{center}
\includegraphics[width=16cm]{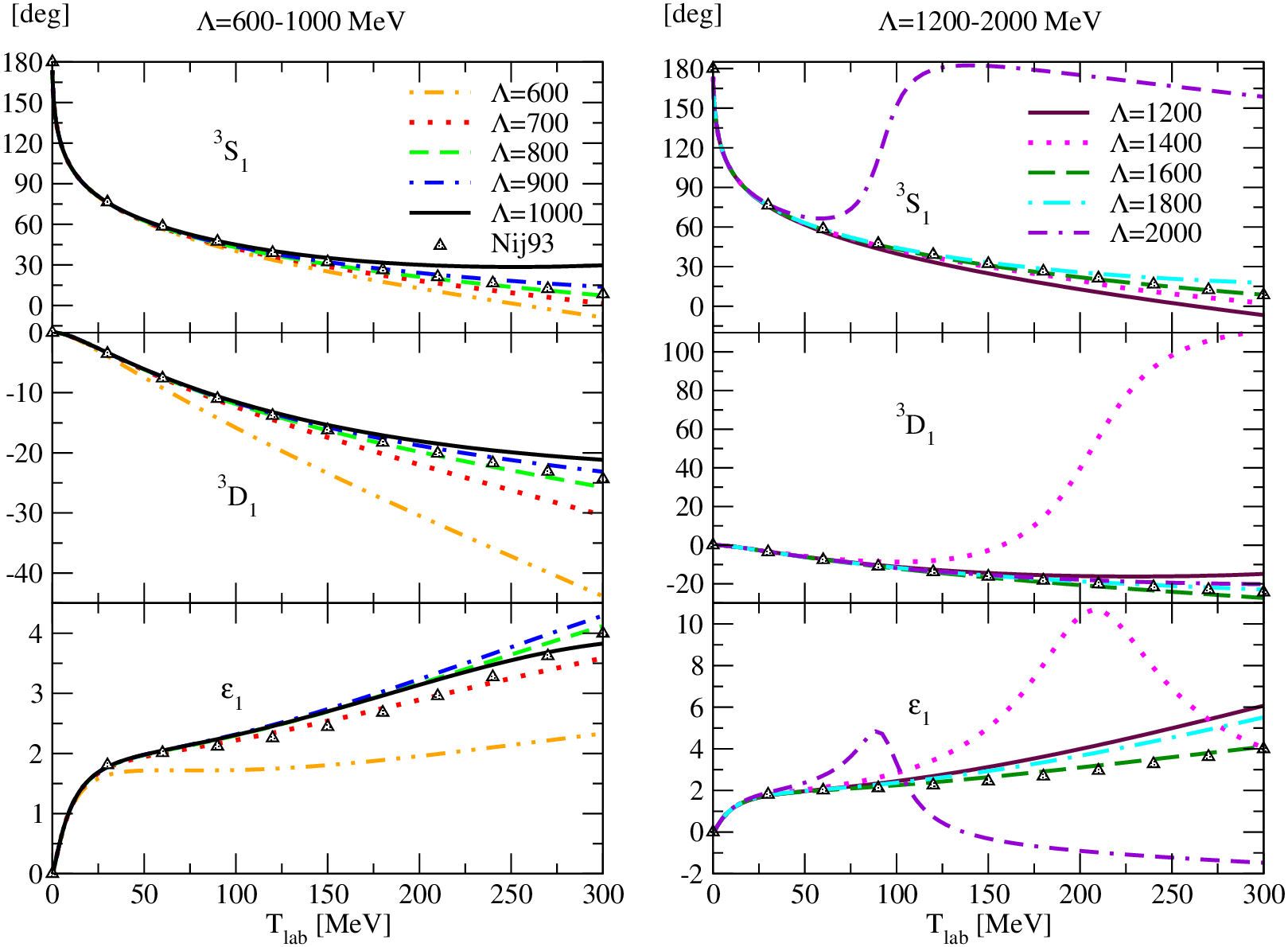}
\end{center}
\par
\vspace{3mm}
\caption{(Color online) The J=1 coupled NN phase shifts as a function of the
laboratory kinetic energy for different cutoffs $\Lambda$ ranging from 0.6
to 2~GeV. The potential employed is DR NNLO with a linear energy dependence
in the central part of the contact term. The results are obtained by three
subtractions with $a_0=5.428$ fm, $\protect\alpha_{20}=2.28\times 10^{-10}$
MeV$^{-3}$ and the phase shift at $T_{lab}=10$ MeV as input. The values of
the Nijmegen phase-shifts~\protect\cite{nnonline} are indicated by the open
triangles.}
\label{fig-fig14}
\end{figure}

The results of using an energy-dependent contact term to renormalize the
three different long-range potentials are shown in Figs.~\ref{fig-fig14}-\ref%
{fig-fulle}. In Fig.~\ref{fig-fig14} we adopt $a_{t}=5.428$ fm, $\alpha
_{20}=2.28\times 10^{-10}$ MeV$^{-3}$~\cite{PVRA05B}, and the Nijmegen PWA
values of $\delta (^{3}S_{1})$, $\delta (^{3}D_{1})$ and $\epsilon _{1}$ at $%
T_{lab}=10$ MeV as our input to renormalize the DR\ NNLO TPE. This choice
does not produce an optimal description of $\epsilon _{1}$ with respect to
the Nijmegen analysis, although it is pretty close.

Actually, $\epsilon_1$ is a notoriously delicate observable, and the choice
made for $\alpha_{20}$ affects it appreciably. Therefore, in the case of DR
NNLO we further adjust $\alpha _{20}$ to $2.25\times 10^{-10}$ MeV$^{-3},$
which yields the best fit of $\epsilon _{1}$ with respect to the Nijmegen
values for $\Lambda =700-1200$ MeV. This result is shown in Fig.~\ref%
{fig-fig15}. Comparison with Fig.~\ref{fig-fig14} shows that this 1.5\%
shift in $\alpha_{20}$ leads to noticeable improvement in $\epsilon_1$ over
this cutoff range, with few other significant changes. The same value, $%
\alpha _{20}=2.25\times 10^{-10}$ MeV$^{-3}$, also gives the best fit for DR
NLO, and this $\alpha_{20}$ produces results for SFR NNLO which coincide
quite well with the Nijmegen data. In addition, $\alpha_{20}=2.25\times
10^{-10}$ MeV$^{-3}$ is still within a few per cent of the value obtained
for this quantity from the potentials constructed in Ref.~\cite{St94}. Such
a difference is certainly within the range of expected N$^3$LO corrections
to our results.

The DR NLO results obtained with $\alpha_{20}=2.25\times 10^{-10}$ MeV$^{-3}$
are shown in Fig.~\ref{fig-fig16}. We see there that the phase shifts
oscillate for $\Lambda \leq 800$ MeV, and then converge for $\Lambda
=900-1800$ MeV. For $\Lambda=2000$ MeV, the phase shifts diverge badly again.

The DR NNLO results present several interesting features. First, there
exists a range of cutoffs $\Lambda =800-900$ MeV where all three phase
shifts agree with the Nijmegen analysis remarkably well\footnote{It is interesting to point out that for $\Lambda=800-900$ MeV, the energy-dependent contact term actually produces phase shifts which fit the Nijmegen analysis better than the momentum-dependent contact term, in both $^1S_0$ and $^3S_1-^3D_1$ channels. This may indicate that this on-shell-energy-dependent contact term represents the short-distance physics quite well in this narrow range of cutoffs.}. However, with
higher cutoffs the phase shifts gradually move away from the Nijmegen
analysis. Second, for $\Lambda =2000(1400) $ MeV, $\delta (^{3}S_{1})(\delta
(^{3}D_{1}))$ diverge and show a resonance-like behavior.

In order to analyze this feature, we calculate the J=1 triplet phase shifts
for a fixed energy of $T_{lab}=10$ and $50$ MeV and a wide range of cutoffs $%
\Lambda =500-5500$ MeV, and plot them in Figs.~\ref{fig-fig18} and \ref%
{fig-fig188}. Both figures show oscillatory behavior of the phase shifts
with respect to $\Lambda $. The S and D-waves have different points of
divergence in this cycle, and (unsurprisingly) affect one another when
either gets large. The oscillation pattern of the mixing angle $\epsilon _{1}
$ is a combination of that appearing in the two phase shifts $\delta
(^{3}S_{1})$ and $\delta (^{3}D_{1})$.

In Fig.~\ref{fig-fulle} we show the results for the SFR NNLO potential with $%
\alpha_{20}=2.25\times 10^{-10}$ MeV$^{-3}$. The phase shifts only show
minor cutoff dependence for $\Lambda <1000$ MeV and converge nicely for
higher $\Lambda$. As far as the fit to the Nijmegen PWA is concerned, a
decent agreement is observed for all $\Lambda $ for $\delta (^{3}S_{1})$ and 
$\delta (^{3}D_{1})$. For $\epsilon _{1}$, the converged curves do not fit
as well as those for $\Lambda =600-800$ MeV. This is not a surprise, since
we  adopted a value of $\alpha _{20}$ which fits $\epsilon _{1}$\ in this
region of cutoff. We found that at higher cutoffs, i.e., $\Lambda\sim 2300$
MeV, the SFR NNLO TPE presents a divergent pattern in the phase shifts.
Similar to the singlet channel, SFR NNLO can defer, but cannot avoid, the
resonance caused by the energy-dependent contact term.

The effective range $r_{0}$ obtained at NLO is $r_0=1.75 \pm 0.01$ fm for $%
900~\mathrm{MeV} \leq \Lambda \leq 1800$ MeV, with the error representing
the extent of the cutoff dependence in that range. For $\Lambda=2000$ MeV, $%
r_0=1.70$fm, which is due to the resonance-like behavior in the phase shifts
created by the energy-dependent contact term. For DR NNLO, $r_0=1.75 \pm 0.01$
fm except for $\Lambda=1400$ and $2000$ MeV, which are just the cutoffs
where the phase shifts also diverge, as shown in Figs.~\ref{fig-fig18} and %
\ref{fig-fig188}. The same behavior in $r_0$ is present in the SFR NNLO,
but for higher cutoffs.

In summary, in the triplet channel the energy-dependent contact term creates
a similar pattern as in the singlet channel. For the DR TPE (SFR NNLO),
there is a highest cutoff $\Lambda =1200 (2300)$ MeV with which the
potential can be iterated in the  LSE before resonances are generated at
energies within the domain of validity of $\chi$ET.

\subsection{Momentum-dependent contact term}

The results of using a momentum-dependent contact term to renormalize the
DR\ NNLO TPE are shown in Fig.~\ref{fig-fig19}. Here we adopt $a_{t}=5.428$
fm and $\alpha _{20}=2.25\times 10^{-10}$ MeV$^{-3}$ as input and then
perform a fit to the phase shift $\delta (^{3}S_{1})$ at $T_{lab}=200$ MeV.
This pins down the three unknown LECs in the potential at this order.

For phase shifts obtained by our two-subtractions-plus-one-fitting method,
once the trial value of $C_2$ is imposed, the other two constants $\lambda $
and $\lambda _{t}$ are fixed by the two pieces of experimental information
that are our input: $\alpha_{20}$ and $a_t$. Thus, the results shown in Fig.~%
\ref{fig-fig19} satisfy the constraints at $T_{lab}=0$ associated with this
information and reproduce the ${}^3$S$_1$ phase shift at  $T_{lab}=200$ MeV.
However, they do not provide the best possible fit to the phase shifts.  In
contradistinction to the previous section, the fit is not appreciably
improved by variations of $\alpha_{20}$. Compared to the results obtained
using the same contact terms in Refs.~\cite{Ep99,EM03,Ep05}, the overall fit
to the Nijmegen phase shifts is noticeably worse. Furthermore, the results
obtained by fixing $\delta (^{3}S_{1})$ at $T_{lab}=200$ MeV do not show
stability for $\delta (^{3}D_{1})$ and $\epsilon _{1}$\ with respect to $%
\Lambda $. This suggests that the TPE\ associated with a momentum-dependent
contact term depends strongly on the choice of renormalization point once $%
\Lambda > 1$ GeV.

We wish to see up to what value of $\Lambda$ it is possible to ameliorate
these difficulties by a different choice of input. Thus we remove the two
constraints imposed by $a_{t}=5.428$ fm and $\alpha _{20}=2.25\times 10^{-10}
$ MeV$^{-3}$ and allow all three NN LECs to vary freely, so that we can
obtain the best overall fit to the phase shifts. The results for DR NLO, DR
NNLO and SFR NNLO are listed in Figs.~\ref{fig-fig21a} and \ref{fig-fig200},
respectively. The fits are a noticeable improvement over Fig.~\ref{fig-fig19}%
. For all three potentials, starting from $\Lambda =700$ MeV, the $^{3}S_{1}$
and $^{3}D_{1}$ phase shifts fit the Nijmegen analysis quite well and have
only minor variation with respect to $\Lambda $ up to $1000$ MeV. A stronger
cutoff dependence shows up only in the mixing parameter $\epsilon_{1}$.
There we observe a clear decrease in the $\Lambda$ sensitivity of the result
as the long-range potential is improved from DR NLO to DR NNLO and then
again to SFR NNLO. We plot $r_{0}$ extracted from those best fits in Fig.~%
\ref{fig-t4}. Again, we see better agreement with the Nijmegen value $%
r_0=1.833$ fm~\cite{PVRA05B} for DR NNLO than for DR NLO. For $700 \le \Lambda \le 1000$ MeV, $r_0$ varies by less than $3\%$ for all three potentials. 

\section{Conclusion}

\label{sec-con}

In this paper we developed subtractive renormalization schemes which can be
applied to NN scattering in $\chi$ET at NNLO and maximize the use of
information from experiment. We showed how to manipulate LSEs containing the
contact interactions of $\chi$ET so as to eliminate constant, tensor, and
energy-dependent contact operators from the equation in favor of pieces of
experimental data. A by-product of our analysis is the result that the
momentum-dependent contact operator cannot be eliminated in this way, and
thus is not straightforwardly related to any NN S-matrix element. We can
still, however, analyze the $\chi$ET potential containing that operator, by
using a mixture of subtractive renormalization and more standard fitting
methods.

The subtractive-renormalization method shows particular advantages in cases
where the LECs undergo limit-cycle or limit-cycle-like behavior, and thus
adjusting several of them to reproduce data represents a difficult fitting
problem. It is also very useful when carrying out calculations at cutoffs $%
\Lambda > 1$ GeV, since  fine-tuning of the unknown NN LECs occurs in that
regime. Finally, the direct input of experimental information allows a
straightforward determination of the impact of the renormalization
point---i.e. the choice of \textit{which} experimental information we
input---on the resulting phase shifts.

Looking first at the case of a constant contact interaction, in the ${}^1$S$%
_0$ channel our results confirm the finding of Ref.~\cite{En08}: when a
constant contact term is adopted, the resulting $^{1}$S$_{0}$ phase shift
converges and becomes independent of $\Lambda$ in the limit $\Lambda
\rightarrow \infty$. This occurs once $\Lambda >1000 (1200)$ for the DR NLO
and SFR NNLO (DR NNLO) potentials. However, we found that in this
calculation the contribution from the NNLO($Q^{3}$) part of the TPE is
larger than the NLO part.

In contrast, in the coupled ${}^3$S$_1-{}^3$D$_1$ channels the dimensionally
regularized interactions cannot be renormalized with a single, constant,
contact interaction. However, when the two-pion exchange is computed using
spectral-function regularization $\Lambda$-independent results are again
obtained at large $\Lambda$: presumably because the short-distance behavior
of this potential is the same as that of one-pion exchange, which is
well-known to be renormalized by a single constant contact term in the ${}^3$%
S$_1-{}^3$S$_1$ part of the potential~\cite{Be02,PVRA04B,Ya08}.

We also considered the contact terms that can occur up to O($Q^2$) in $\chi$%
ET, examining both energy-dependent and momentum-dependent contact terms.
For both the ${}^1$S$_0$ and ${}^3$S$_1-{}^3$D$_1$ channels, the phase
shifts obtained from dimensionally regularized two-pion exchange show
oscillatory behavior as a function of $\Lambda$ when energy-dependent
contact terms are used. This is associated with the appearance, and
subsequent movement with energy, of poles in the NN amplitude. The first
cutoff at which such a pole appears in the domain of validity of $\chi$ET is
around $\Lambda=1000$ MeV.

We also found that cutoffs less than about $600$ MeV sometimes produced
resonant features in the phase shifts. We attribute this failure to the
propensity of these cutoffs to eliminate too much of the central attraction
in the TPE that is responsible for reproducing the energy dependence of the
S-wave phase shifts extracted from experiment---especially in the ${}^1$S$_0$%
.

In between these two limits, i.e. for 600 MeV $< \Lambda <$ 1 GeV, we found
that the 1993 Nijmegen PWA phase shifts, in both the ${}^1$S$_0$ and ${}^3$S$%
_1-{}^3$D$_1$ channels, could be quite well reproduced using a
short-distance potential that includes an energy-dependent contact term,
together with a long-distance potential calculated to $O(Q^3)$ in $\chi$PT.
With an energy-dependent contact term good fits to the Nijmegen phase shifts
can be obtained at cutoffs above 1 GeV, but only in fairly narrow windows of 
$\Lambda$, where the energy of the spurious bound state produced by the
energy-dependent contact term is large compared to the energy scales of
interest.

Finally, if the spectral-function regularized TPE is adopted, then the range
of the cutoff one can use before the resonance-like behavior occurs is
greatly increased.

Turning to the case of the momentum-dependent contact interaction: we found
that in the case of this type of contact interaction there is a significant
improvement with respect to the Nijmegen phase shift when one improves the
long-range part of the potential from NLO to NNLO.  Indeed, if the $\chi$ET
NLO potential is employed there are certain observables (e.g. the effective
range) that \textit{cannot} be reproduced if we restrict ourself to real
values of the LECs in the NN potential. If we wish our NLO calculation to
reproduce the effective range in the ${}^1$S$_0$ we are restricted to $%
\Lambda < 600$~MeV. This problem is absent at NNLO.

We examined the renormalization-point dependence of our results, and found
that in the ${}^3$S$_1-{}^3$D$_1$ channel at $\Lambda=800$ MeV the results
from both the DR and SFR TPE up to NNLO are almost independent of the
renormalization point. However, once $\Lambda =1000$ MeV is reached, the
results with the DR NNLO potential show appreciable dependence on the choice
of experimental input. Once again, the SFR TPE is better behaved in this
regard: there is only minor renormalization-point dependence for the SFR
NNLO potential from $\Lambda=600$ MeV until $\Lambda=1800$ MeV, and the
renormalization-point independence breaks down at $\Lambda=2000$ MeV.

The use of a momentum-dependent potential and the expressions for the NNLO
TPE should allow us to confirm previous results for the triplet channel
(e.g. those of Ref.~\cite{Ep99}). However, we found that if we impose strict
relations between the three unknown LECs in order to reproduce particular
pieces of experimental data (e.g. $a_{t}$ and $\alpha _{20}$), the resulting
fit to the Nijmegen phase shifts is in general worse than that obtained by
direct fitting of data over a finite range of energies~\cite{Ep99}. Only
when all the three constants are allowed to vary independently do we get
best fits, which are comparable in quality to the results in~\cite{Ep99,
PVRA06A}. However, we want be emphasize that this fit can only be obtained
for cutoffs $\Lambda =700$--$900$ MeV for the DR NNLO TPE. (For the SFR NNLO
TPE, it is possible to obtain a reasonable fit up to $\Lambda=2000$ MeV.)
There is thus an appreciable difference between using two subtractions that
fix the zero-energy behavior of the amplitude and performing a best fit
involving all three $J=1$, $S=1$ NN LECs. That difference is an indication
that the triplet S-wave phase shifts obtained from chiral TPE depend on the
renormalization point rather strongly, despite the fact that they are quite
cutoff independent in this window of $\Lambda$.

In summary, we have probed the renormalization-point and cutoff dependence
of chiral potentials which include TPE when these potentials are iterated
using the LSE. We find that energy-dependent contact terms---which might be
considered useful as they allow one to evade certain theorems regarding
short-distance potentials~\cite{Wi55,PC96}---tend to introduce unphysical
resonances in the results once $\Lambda >1$ GeV. These resonances are
generically absent when a momentum-dependent contact term is employed. Since
the $\chi $ET power counting suggests that energy- and momentum dependent
contact interactions are equivalent up to the order to which we work this
implies that the power counting is only a useful guide for $\Lambda <1$ GeV.

Although the momentum-dependent contact interaction does not produce the
dramatic failures to agree with experiment that occur in the
energy-dependent case, it does have some difficulties. Chief among these is
the failure of certain aspects of the NLO calculation. Serious doubt must
exist regarding the convergence of an expansion where (at least for the ${}^1$S$_0$
channel) LO is in poor agreement with experiment, NLO cannot reproduce the
effective range, and it is not until NNLO that anything like a reasonable
picture of the phase shifts emerges.

The less singular interactions obtained with spectral-function
regularization avoid many of the difficulties inherent in the use of their
dimensionally regularized counterparts. The cutoff dependence is almost
always weaker, with the cutoff dependence of phase shifts found with the SFR
NNLO potential and a constant contact term disappearing altogether for $%
\Lambda > 1$ GeV. However, $\epsilon_1$ cannot be reproduced in this range
of cutoffs, and the issue with the effective range at NLO persists.

We conclude that it makes little sense to iterate the two-pion-exchange
potentials obtained from chiral perturbation theory using the
Lippmann-Schwinger equation unless one restricts the momentum in the LSE
integral to lie below $\sim 1$ GeV~\cite{Le97,EM06,Dj07}. In this domain,
and for the case of a momentum-dependent contact term, our
subtractive-renormalization method produces results which agree with
previous analyses~\cite{PVRA06A, PVRA06B,Ep99, epsfr, En08}.


\vfill


\begin{acknowledgments}
This work was performed in part under the auspices of the U.~S. Department
of Energy, Office of Nuclear Physics, under contract No. DE-FG02-93ER40756
with Ohio University. We thank the Ohio Supercomputer Center (OSC) for the
use of their facilities under grant PHS206. D.R.P. is grateful for financial
support form the Mercator programme of the Deutsche Forschungsgemeinschaft,
for the hospitality of the Theoretical Physics group at the University of
Manchester during the initial stages of this work, and for useful
conversations with Evgeny Epelbaum.
\end{acknowledgments}




\appendix

\section{Generating $t(E^{\ast })$ from a given value of $\protect\delta %
(E^{\ast })$}

\label{appendixa}

For any given energy $E^\ast$ the on-shell value of the $t$ matrix is
related to the phase shift $\delta (E^{\ast })$ by 
\begin{eqnarray}
t(p_{0}^{\ast },p_{0}^{\ast },E^{\ast })=\frac{f_{00}(E^\ast)}{-Mp_{0}^{\ast
}}= \frac{e^{i\delta (E^{\ast })}\sin \delta (E^{\ast })}{-Mp_{0}^{\ast }},
\label{eq:4.10}
\end{eqnarray}
where $f_{00}(E^\ast)$ represents the scattering amplitude and $p_{0}^{\ast
}= \sqrt{ME^{\ast }}$. Then we write the half-shell and on-shell LS
equations as 
\begin{eqnarray}
t(p^{\ast ^{\prime }},p_{0}^{\ast };E^{\ast }) &=&\lambda +\gamma E^{\ast
}+v_{LR }(p^{\ast ^{\prime }},p_{0}^{\ast })+\frac{2}{\pi }%
M\int_{0}^{\Lambda }dp^{\ast }\;p^{\ast }{}^{2}\left( \frac{\lambda +\gamma
E^{\ast }+v_{LR }(p^{\ast ^{\prime }},p^{\ast })}{p_{0}^{\ast 2}-p^{\ast
}{}^{2}+i\varepsilon }\right) \;t(p^{\ast },p_{0}^{\ast };E^{\ast }),
\label{eq:4.11} \\
t(p_{0}^{\ast },p_{0}^{\ast };E^{\ast }) &=&\lambda +\gamma E^{\ast }+v_{LR
}(p_{0}^{\ast },p_{0}^{\ast })+\frac{2}{\pi }M\int_{0}^{\Lambda }dp^{\ast
}\;p^{\ast }{}^{2}\left( \frac{\lambda +\gamma E^{\ast }+v_{LR }(p_{0}^{\ast
},p^{\ast })}{p_{0}^{\ast 2}-p^{\ast }{}^{2}+i\varepsilon } \right)
\;t(p^{\ast },p_{0}^{\ast };E^{\ast }).  \label{eq:4.12}
\end{eqnarray}
Subtracting Eq.~(\ref{eq:4.12}) from Eq.~(\ref{eq:4.13}) leads to 
\begin{eqnarray}
t(p^{\ast \prime },p_{0}^{\ast };E^{\ast })&=&v_{LR }(p^{\ast \prime
},p_{0}^{\ast })-v_{LR }(p_{0}^{\ast },p_{0}^{\ast })+\frac{e^{i\delta
(E^{\ast })}\sin \delta (E^{\ast })}{-Mp_{0}^{\ast }} \cr &+& \frac{2}{\pi }
M\; \int_{0}^{\Lambda }dp^{\ast }\;p^{\ast }{}^{2}\;\left( \frac{v_{LR
}(p^{\ast \prime },p^{\ast })-v_{LR }(p_{0}^{\ast },p^{\ast })}{ p_{0}^{\ast
2}-p^{\ast }{}^{2}+i\varepsilon }\right) \;t(p^{\ast },p_{0}^{\ast };E^{\ast
}).  \label{eq:4.13}
\end{eqnarray}
The above equation does not contain any unknown quantities and can solved
with standard techniques. With $t(p^{\ast \prime },p_{0}^{\ast };E^{\ast })$
at hand, we can carry out another subtraction to get $t(p^{\ast \prime
},k^{\ast };E^{\ast })$, where $k^{\ast }$ represents any arbitrary
momentum. 
\begin{eqnarray}
t(p^{\ast \prime },k^{\ast };E^{\ast })=\lambda +\gamma E^{\ast }+v_{LR
}(p^{\ast \prime },k^{\ast })+\frac{2}{\pi }M\;\int_{0}^{\Lambda }dp^{\ast
}\;p^{\ast }{}^{2}\;\left( \frac{\lambda +\gamma E^{\ast }+v_{LR }(p^{\ast
\prime },p^{\ast })}{p_{0}^{\ast 2}-p^{\ast }{}^{2}+i\varepsilon }\right)
\;t(p^{\ast },k^{\ast };E^{\ast }),  \label{eq:4.14}
\end{eqnarray}
\begin{eqnarray}
t(p_{0}^{\ast },k^{\ast };E^{\ast })=\lambda +\gamma E^{\ast }+v_{LR
}(p_{0}^{\ast },k^{\ast })+\frac{2}{\pi }M\;\int_{0}^{\Lambda }dp^{\ast
}\;p^{\ast }{}^{2}\;\left( \frac{\lambda +\gamma E^{\ast }+v_{LR
}(p_{0}^{\ast },p^{\ast })}{p_{0}^{\ast 2}-p^{\ast }{}^{2}+i\varepsilon }%
\right) \;t(p^{\ast },k^{\ast };E^{\ast }).  \label{eq:4.15}
\end{eqnarray}
The difference of Eq. (\ref{eq:4.14}) and Eq. (\ref{eq:4.15}) gives 
\begin{eqnarray}  \label{eq:4.16}
t(p^{\ast \prime },k^{\ast };E^{\ast })-t(p_{0}^{\ast },k^{\ast };E^{\ast
})&=& v_{LR }(p^{\ast \prime },k^{\ast })-v_{LR }(p_{0}^{\ast },k^{\ast }) %
\cr &+& \frac{2}{\pi }M\;\int_{0}^{\Lambda }dp^{*}\;p^{* }{}^{2}\;\left( 
\frac{v_{LR }(p^{\ast \prime },p^{\ast })-v_{LR }(p_{0}^{\ast },p^{\ast })}{%
p_{0}^{\ast 2}-p^{\ast }{}^{2}+i\varepsilon } \right) t(p^{\ast },k^{\ast
};E^{\ast }).  \notag \\
\end{eqnarray}
Using $t(p_{0}^{\ast },k^{\ast };E^{\ast })=t(k^{\ast },p_{0}^{\ast
};E^{\ast })$ we can calculate $t(p^{\ast \prime },k^{\ast };E^{\ast })$ by
solving Eq. (\ref{eq:4.16}).

\section{Details of the subtraction method in the J=1 triplet channel}

\label{appendixb}

Let us start with the on-shell and half-off-shell LSE in this, two-channel,
case. We will consider a momentum-dependent short-distance potential, so we
have: 
\begin{eqnarray}  \label{eq:11}
\left( 
\begin{array}{cc}
a_{t}/M & 0 \\ 
0 & 0%
\end{array}
\right) &=& \left( 
\begin{array}{cc}
\lambda +v_{LR }^{00}(0,0) & 0 \\ 
0 & 0%
\end{array}
\right)  \notag \\
&-&\frac{2}{\pi }M\;\int_{0}^{\Lambda }dp^{\prime }\; \left( 
\begin{array}{cc}
\lambda +C_2 p^{\prime 2}+v_{LR }^{00}(0,p^{\prime }) & \lambda
_{t}p^{\prime 2}+v_{LR }^{02}(0,p^{\prime }) \\ 
0 & 0%
\end{array}
\right) \;\left( 
\begin{array}{cc}
t_{00}(p^{\prime },0;0) & 0 \\ 
t_{20}(p^{\prime },0;0) & 0%
\end{array}
\right)  \notag \\
\end{eqnarray}
\begin{eqnarray}  \label{eq:12}
\left( 
\begin{array}{cc}
t_{00}(p,0;0) & 0 \nonumber \\ 
t_{20}(p,0;0) & 0%
\end{array}
\right) &= &\left( 
\begin{array}{cc}
\lambda +C_2 p^{2}+v_{LR }^{00}(p,0) & 0 \\ 
\lambda _{t}p^{2}+v_{LR }^{20}(p,0) & 0%
\end{array}
\right)  \notag \\
&-&\frac{2}{\pi }M\;\int_{0}^{\Lambda }dp^{\prime } \left( 
\begin{array}{cc}
\lambda +C_2 (p^{2}+p^{\prime 2})+v_{LR }^{00}(p,p^{\prime }) & \lambda
_{t}p^{\prime 2}+v_{LR }^{02}(p,p^{\prime }) \\ 
\lambda _{t}p^{2}+v_{LR }^{20}(p,p^{\prime }) & v_{LR }^{22}(p,p^{\prime })%
\end{array}
\right) \;\left( 
\begin{array}{cc}
t_{00}(p^{\prime },0;0) & 0 \\ 
t_{20}(p^{\prime },0;0) & 0%
\end{array}
\right)  \notag .\\
\end{eqnarray}

\noindent Subtracting Eq.~(\ref{eq:11}) from Eq.~(\ref{eq:12}) gives 
\begin{eqnarray}
\left( 
\begin{array}{cc}
t_{00}(p,0;0)-a_{t}/M & 0 \\ 
t_{20}(p,0;0) & 0%
\end{array}
\right) &= & \left( 
\begin{array}{cc}
C_2 p^{2}+v_{LR }^{00}(p,0) & 0 \\ 
\lambda _{t}p^{2}+v_{LR }^{20}(p,0) & 0%
\end{array}
\right)  \notag \\
&-&\frac{2}{\pi }M\;\int_{0}^{\Lambda }dp^{\prime } \left( 
\begin{array}{cc}
C_2 p^{2}+v_{LR }^{00}(p,p^{\prime })-v_{LR }^{00}(0,p^{\prime }) & v_{LR
}^{02}(p,p^{\prime })-v_{LR }^{02}(0,p^{\prime })\nonumber \\ 
\lambda _{t}p^{2}+v_{LR }^{20}(p,p^{\prime }) & v_{LR }^{22}(p,p^{\prime })%
\end{array}
\right) \\
&& \qquad \qquad \times \left( 
\begin{array}{cc}
t_{00}(p^{\prime },0;0) & 0 \\ 
t_{20}(p^{\prime },0;0) & 0%
\end{array}
\right).  \label{eq:13}
\end{eqnarray}

\noindent The only unknown in Eq.~(\ref{eq:13}) is $\lambda _{t}$, thus we
can express 
\begin{equation}
\lambda _{t}=\frac{t_{20}(p,0;0)-v_{LR }^{20}(p,0)+\frac{2}{\pi }
M\;\int_{0}^{\Lambda }dp^{\prime }\;\left( v_{LR }^{20}(p,p^{\prime
})t_{00}(p^{\prime },0;0)+v_{LR }^{22}(p,p^{\prime })t_{20}(p^{\prime
},0;0)\right) }{p^{2}(1-\frac{2}{\pi }M\;\int_{0}^{\Lambda }dp^{\prime
}t_{00}(p^{\prime },0;0))}.  \label{eq:lambdat2}
\end{equation}

Similarly, once $\lambda _{t}$ is calculated, from Eq.~(\ref{eq:11}) we can
express $\lambda $ as 
\begin{equation}  \label{eq:15}
\lambda =\frac{a_{t}/M-v_{LR }^{00}(0,0)+\frac{2}{\pi }M\;\int_{0}^{\Lambda
}dp^{\prime }\;[\left( C_2 p^{\prime 2}+v_{LR }^{00}(0,p^{\prime
}))t_{00}(p^{\prime },0;0)+(\lambda _{t}p^{\prime 2}+v_{LR
}^{02}(0,p^{\prime }))t_{20}(p^{\prime },0;0)\right) ]}{1-\frac{2}{\pi }%
M\;\int_{0}^{\Lambda }dp^{\prime }t_{00}(p^{\prime },0;0)}.
\end{equation}

\noindent Thus, by using the two scattering length $a_{t}$ and $\alpha _{20}$
together with a trial value for $C_2 $, we can obtain the other two unknown
constants $\lambda $ and $\lambda _{t}$. This determines the complete NN
potential at this order, and it is then a straightforward matter to compute
the phase shifts using the LSE.

The LSE with the energy-dependent contact term (case B) satisfies exactly
the same relation as Eq.~(\ref{eq:9}) and Eq.~(\ref{eq:10}), while Eq.~(\ref%
{eq:7.6}) is replaced by 
\begin{eqnarray}  \label{eq:16}
t_{00}(p,0;0)&-&\frac{a_{t}}{M}=v_{LR }^{00}(p,0)-v_{LR }^{00}(0,0)  \notag
\\
&-&\frac{2}{\pi }M\int_{0}^{\Lambda }dp^{\prime } \Big[ \left( v_{LR
}^{00}(p,p^{\prime })-v_{LR }^{00}(0,p^{\prime })\right) \ t_{00}(p^{\prime
},0;0) + \left( v_{LR }^{02}(p,p^{\prime })-v_{LR }^{02}(0,p^{\prime })
\right) t_{20}(p^{\prime },0;0) \Big]  \notag .\\
\end{eqnarray}

\noindent Equation~(\ref{eq:16}), together with Eq.~(\ref{eq:9}), can be
used to solve for $t(p,0;0) $. Note that here we don't need to input a trial
value for  $C_2 $. Once we have $t_{00}(p,0;0)$ at hand, we can substitute
it, together with the $t_{20}$ obtained from Eq.~(\ref{eq:10}), back into
Eq.~(\ref{eq:7.5}) (which is unaffected by the issue of energy- vs.
momentum-dependence and so is the same as Eq.~(\ref{eq:lambdat2})) and
obtain $\lambda _{t}$.

The value of $\lambda $ then can be obtained from the new version of Eq.~(%
\ref{eq:11}), 
\begin{eqnarray}  \label{eq:19}
\left( 
\begin{array}{cc}
a_{t}/M & 0 \\ 
0 & 0%
\end{array}
\right) =\left( 
\begin{array}{cc}
\lambda +v_{LR }^{00}(0,0) & 0 \\ 
0 & 0%
\end{array}
\right) -\frac{2}{\pi }M\;\int_{0}^{\Lambda }dp^{\prime } \left( 
\begin{array}{cc}
\lambda +v_{LR }^{00}(0,p^{\prime }) & \lambda _{t}p^{\prime 2}+v_{LR
}^{02}(0,p^{\prime }) \\ 
0 & 0%
\end{array}%
\right) \left( 
\begin{array}{cc}
t_{00}(p^{\prime },0;0) & 0 \\ 
t_{20}(p^{\prime },0;0) & 0%
\end{array}
\right),  \notag \\
\end{eqnarray}
which gives 
\begin{equation}  \label{eq:20}
\lambda =\frac{a_{t}/M-v_{LR }^{00}(0,0)+\frac{2}{\pi }M\;\int_{0}^{\Lambda
}dp^{\prime }\;[v_{LR }^{00}(0,p^{\prime })t_{00}(p^{\prime },0;0)+(\lambda
_{t}p^{\prime 2}+v_{LR }^{02}(0,p^{\prime }))t_{20}(p^{\prime },0;0)]}{1-%
\frac{2}{\pi }M\;\int_{0}^{\Lambda }dp^{\prime }t_{00}(p^{\prime },0;0)}.
\end{equation}

Finally, in order to solve for $\gamma$, one needs a third piece of
information as input. A convenient choice is the phase shifts at an energy $E
$. From $^{3}S_{1}$, $^{3}D_{1}$ and $\epsilon _{1}$ at a specific energy $E$
we compute $t(p_{0},p_{0};E)$. Then, from a modified Eq.~(\ref{eq:7}), 
\begin{eqnarray}  \label{eq:21}
t(p,p_{0};E)&-&t(p_{0},p_{0};E)=\left( 
\begin{array}{cc}
v_{LR }^{00}(p,p_{0})-v_{LR }^{00}(p_{0},p_{0}) & v_{LR
}^{02}(p,p_{0})-v_{LR }^{02}(p_{0},p_{0}) \nonumber \\ 
\lambda _{t}(p^{2}-p_{0}^{2})+v_{LR }^{20}(p,p_{0})-v_{LR }^{20}(p_{0},p_{0})
& v_{LR }^{22}(p,p_{0})-v_{LR }^{22}(p_{0},p_{0})%
\end{array}
\right)  \notag \\
+\frac{2}{\pi }M\int_{0}^{\Lambda } &dp^{\prime }& \frac{p{}^{\prime 2}}{%
p_{0}^{2}-p^{\prime }{}^{2}+i\varepsilon } \left( 
\begin{array}{cc}
v_{LR }^{00}(p,p^{\prime })-v_{LR }^{00}(p_{0},p^{\prime }) & v_{LR
}^{02}(p,p^{\prime })-v_{LR }^{02}(p_{0},p^{\prime }) \\ 
\lambda _{t}(p^{2}-p_{0}^{2})+v_{LR }^{20}(p,p^{\prime })-v_{LR
}^{20}(p_{0},p^{\prime }) & v_{LR }^{22}(p,p^{\prime })-v_{LR
}^{22}(p_{0},p^{\prime })%
\end{array}
\right) \ t(p^{\prime },p_{0};E)  \notag ,
\end{eqnarray}

\noindent we solve for $t(p,p_{0};E)$. Once we have $t(p,p_{0};E)$ at hand, $%
\gamma$ can be obtained from a modified Eq.~(\ref{eq:6}): 
\begin{eqnarray}  \label{eq:21.1}
\left( 
\begin{array}{cc}
t_{00}(p,p_{0};E) & t_{02}(p,p_{0};E) \\ 
t_{20}(p,p_{0};E) & t_{22}(p,p_{0};E)%
\end{array}
\right) &= &\left( 
\begin{array}{cc}
\lambda +\gamma E+v_{LR }^{00}(p,p_{0}) & \lambda _{t}p_{0}^{2}+v_{LR
}^{02}(p,p_{0}) \\ 
\lambda _{t}p^{2}+v_{LR }^{20}(p,p_{0}) & v_{LR }^{22}(p,p_{0})%
\end{array}
\right)  \notag \\
+\frac{2}{\pi }M\int_{0}^{\Lambda } &dp^{\prime } & \frac{p{}^{\prime 2}}{%
p_{0}^{2}-p^{\prime }{}^{2}+i\varepsilon } \left( 
\begin{array}{cc}
\lambda +\gamma E+v_{LR }^{00}(p,p^{\prime }) & \lambda _{t}p^{\prime
2}+v_{LR }^{02}(p,p^{\prime }) \\ 
\lambda _{t}p^{2}+v_{LR }^{20}(p,p^{\prime }) & v_{LR }^{22}(p,p^{\prime })%
\end{array}
\right)  \notag \\
&\times & \left( 
\begin{array}{cc}
t_{00}(p^{\prime },p_{0};E) & t_{02}(p^{\prime },p_{0};E) \\ 
t_{20}(p^{\prime },p_{0};E) & t_{22}(p^{\prime },p_{0};E),%
\end{array}
\right),
\end{eqnarray}
as: 
\begin{equation}  \label{eq:22}
\gamma =-\frac{\lambda -t_{00}(p,p_{0};E)+v_{LR }^{00}(p,p_{0})+\frac{2} {%
\pi }M\int_{0}^{\Lambda }dp^{\prime }p{}^{\prime 2}\frac{[\lambda +v_{LR
}^{00}(p,p^{\prime })]t_{00}(p^{\prime },p_{0};E)+[\lambda _{t}p{}^{\prime
2}+v_{LR }^{02}(p,p^{\prime })]t_{20}(p^{\prime },p_{0};E)} {%
p_{0}^{2}-p^{\prime }{}^{2}+i\varepsilon }}{E[1+\frac{2}{\pi }
M\;\int_{0}^{\Lambda }dp^{\prime }p{}^{\prime 2}\frac{t_{00}(p^{\prime
},p_{0};E)}{p_{0}^{2}-p^{\prime }{}^{2}+i\varepsilon }]}.
\end{equation}

\noindent Note that here we have chosen to solve $\gamma $ by matching the $%
l=l^{\prime }=0$ part of the matrix element in both sides of Eq.~(\ref%
{eq:21.1}). One also could have chosen to solve $\gamma $ by matching the $%
ll^{\prime}=02$ or the $ll^{\prime}=22$ part of the equation. This
corresponds to the fact that we have three phase shifts in the coupled
triplet, namely $^{3}$S$_{1}$, $^{3}$D$_{1}$ and $\varepsilon _{1}$, however
we only need one additional condition to determine the value of the third
unknown $\gamma$.


\section{Two potential analysis of the transition matrix and its relation to
bound states}

\label{appendixc}

Here we want to consider the TPE potential with an energy dependent
contact term, i.e., 
\begin{eqnarray}
\lambda +v_{LR }(p^{\prime },p) +\gamma E \equiv V_{1}+V_{2}.  \label{eq:5.3}
\end{eqnarray}
where $V_2$ should contain the energy dependence. The general derivation of
solution of the LS equation based on the Gellman-Goldberger relation~\cite%
{rmthaler} leads to the following expression for the $t$-matrix 
\begin{eqnarray}
t=t_{1}+\left( 1+t_{1}g_{0}\right )V_{2} \left(1+g_{0}t \right).
\label{eq:5.8}
\end{eqnarray}
Eq. (\ref{eq:5.8}) can be solved by iteration, and the first order 
\begin{eqnarray}
t^{(1)} = t_{1}+\left( 1+t_{1}g_{0}\right )V_{2} \left(1+g_{0}t_1 \right)
\label{eq:5.8a}
\end{eqnarray}
gives the well known DWBA expression. However, we do not want to follow
along that line. Instead we will use Eq.~(\ref{eq:5.8a}) to derive a series
representation of $t$ in terms of $V_2$ and easily calculable integrals.
Explicitly the result up to first order in $V_2$, which we denote by $t^{(1)}
$, is given as 
\begin{eqnarray}
t^{(1)}(p,p^{\prime },E) &= & t_{1}(p,p^{\prime };E)+V_{2}(E)+\frac{2}{\pi }
\int_{0}^{\Lambda }dp^{\ast }\;p^{\ast }{}^{2}\frac{t_{1}(p,p^{\ast
};E)V_{2}(E)} {E-\frac{p^{\ast 2}}{M}+i\epsilon }+\frac{2}{\pi }
\int_{0}^{\Lambda }dp^{\ast }\;p^{\ast }{}^{2}\frac{V_{2}(E)t_{1}(p^{\ast
},p^{\prime };E)}{E-\frac{p^{\ast 2}}{M}+i\epsilon } \cr &+& (\frac{2}{\pi })^2
\int_{0}^{\Lambda }dp^{\ast }\;p^{\ast }{}^{2}dp^{^{\prime \prime }}\;
p^{^{\prime \prime }}{}^{2}\frac{t_{1}(p,p^{\ast };E)V_{2}(E)
t_{1}(p^{^{\prime \prime }},p^{\prime };E)}{(E-\frac{p^{\ast 2}}{ M}%
+i\epsilon ) (E-\frac{p^{^{\prime \prime }2}}{M}+i\epsilon )}.
\label{eq:5.9}
\end{eqnarray}
Since $V_{2} = \gamma E$ does not depend on the momenta $p$ and $p\prime $
it can always be taken out of the integral. Therefore, we define 
\begin{eqnarray}
\Gamma (p;E) \equiv\frac{2}{\pi }\int_{0}^{\Lambda }dp^{\ast }\;p^{\ast
}{}^{2} \frac{t_{1}(p,p^{\ast };E)}{E-\frac{p^{\ast 2}}{M}+i\epsilon },
\label{eq:5.10}
\end{eqnarray}
and observe that we may always make the replacement $t_1 G_0 \rightarrow
\Gamma$ and $G_0 t_1 \rightarrow \Gamma$ if the operators in question appear
next to a $V_2$. Using this notation we can rewrite Eq. (\ref{eq:5.9}) as 
\begin{eqnarray}
t^{(1)}=t_{1}+V_{2}+\Gamma V_{2}+V_{2}\Gamma +\Gamma V_{2}\Gamma .
\label{eq:5.11}
\end{eqnarray}
In the same notation, the sum up to second order in $V_2$, $t^{(2)}$, will
be 
\begin{eqnarray}
t^{(2)}&=& t_{1}+(1+t_{1}g_{0})V_{2}+(1+t_{1}g_{0})V_{2}g_{0}t^{(1)} \cr %
&=&t_{1}+V_{2}+\Gamma V_{2}+(1+t_{1}g_{0})V_{2}g_{0}t^{(1)} \cr &=&
t_{1}+V_{2}+\Gamma V_{2}+V_{2}\Gamma +V_{2}g_{0}V_{2}+V_{2}g_{0}\Gamma
V_{2}+V_{2}g_{0}V_{2}\Gamma +V_{2}g_{0}t_{1}g_{0}V_{2}\Gamma \cr & & +
t_{1}g_{0}\left[ V_{2}\Gamma +V_{2}g_{0}V_{2}+V_{2}g_{0}\Gamma
V_{2}+V_{2}g_{0}V_{2}\Gamma +V_{2}g_{0}t_{1}g_{0}V_{2}\Gamma \right] .
\end{eqnarray}
Let us further define 
\begin{eqnarray}
f(E) &\equiv & (\frac{2}{\pi })^{2}\int_{0}^{ \Lambda }dp^{\ast
}dp\;p^{\ast }{}^{2}p^{2} \frac{t_{1}(p,p^{\ast };E)}{(E- \frac{p^{2}}{M}%
+i\epsilon ) (E-\frac{p^{\ast 2}}{M}+i\epsilon )}, \cr G(E)&\equiv & \frac{2}{%
\pi }\int_{0}^{\Lambda }dp\;\frac{p^{2}}{(E- \frac{ p^{2}}{M}+i\epsilon )},
\label{eq:5.13}
\end{eqnarray}
and note that, when evaluated between two $V_2$'s, $g_0 t_1 g_0$ may be
replaced by $f(E)$. Consequently $t^{(1)}$ and $t^{(2)}$ can be simplified
as 
\begin{eqnarray}
t^{(1)}&=&t_{1}+(1+\Gamma )(V_{2})(1+\Gamma ) \cr t^{(2)}&=&t_{1}+(1+\Gamma
)(V_{2}+V_{2}[G+f]V_{2})(1+\Gamma ),  \label{eq:5.15}
\end{eqnarray}
The next order, $t^{(3)}$ reads 
\begin{eqnarray}
t^{(3)}=t_{1}+(1+\Gamma )\left[
V_{2}+V_{2}[G+f]V_{2}+V_{2}[G+f]^{2}V_{2}^{2} \right] (1+\Gamma ).
\label{eq:5.16}
\end{eqnarray}
Continuing, the N$^{th}$ order $t^{(N)}$ can be written as 
\begin{eqnarray}
t^{(N)}=t_{1}+(1+\Gamma )\left[ \sum_{l=1}^N [V_{2}^{l}(f+G)^{l-1}]\right]
(1+\Gamma ).  \label{eq:5.17}
\end{eqnarray}
Letting $N\rightarrow \infty $, we obtain the formal sum 
\begin{eqnarray}
t=t_{1}+(1+\Gamma )\left[ \frac{1}{\frac{1}{V_{2}}-f-G}\right] (1+\Gamma ).
\label{eq:5.18}
\end{eqnarray}
This tells us that $t-$matrix diverges when $\frac{1}{V_{2}} =f(E)+G(E)$. We
have performed such a calculation for the bound states discussed in Sec.~\ref%
{sec-results1S0endep} and found that Eq. (\ref{eq:5.18}) indeed holds at the
energies where they occur.



\clearpage




\begin{table}[tbp]
\begin{tabular}{|l|cccc|}
\hline
& NijmII~\cite{St94} & Reid93~\cite{St94} & CDBonn~\cite{CDBONN} & Nij93~%
\cite{St94} \\ \hline\hline
$\alpha_{20}$ & $2.28$ & $2.28$ & $2.09$ & $2.18$ \\ \hline\hline
\end{tabular}%
\caption{Value of the generalized scattering length $\protect\alpha_{20}$
extracted from various potentials, in units of $10^{-10}$ MeV$^{-3}$. }
\label{table-5}
\end{table}

\clearpage


\noindent

\vspace{15mm}

\newpage

\vspace{15mm}

\clearpage

\begin{figure}[tbp]
\begin{center}
\includegraphics[width=16cm]{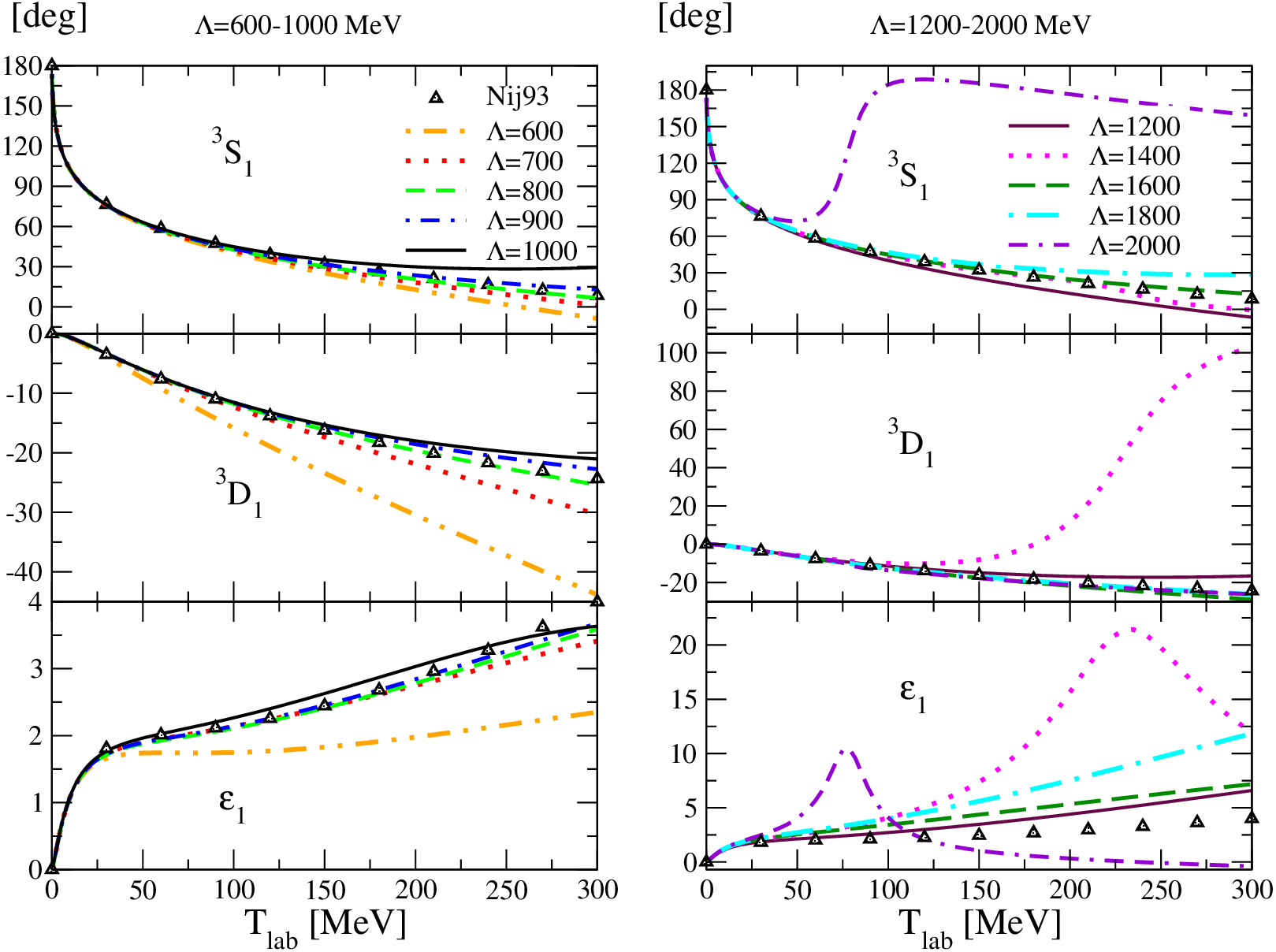}
\end{center}
\par
\vspace{3mm}
\caption{(Color online) The J=1 coupled NN phase shifts as a function of the
laboratory kinetic energy for different cutoffs $\Lambda$ ranging from 0.6
to 2~GeV. The potential employed is the DR NNLO with a linear energy
dependence in the central part of the contact term. The results are obtained
by three subtractions with $a_0=5.428$ fm, $\protect\alpha_{20}=2.25\times
10^{-10}$ MeV$^{-3}$ and the phase shift at $T_{lab}=10$ MeV as input. The
values of the Nijmegen phase-shifts~\protect\cite{nnonline} are indicated by
the open triangles.}
\label{fig-fig15}
\end{figure}

\begin{figure}[tbp]
\begin{center}
\includegraphics[width=16cm]{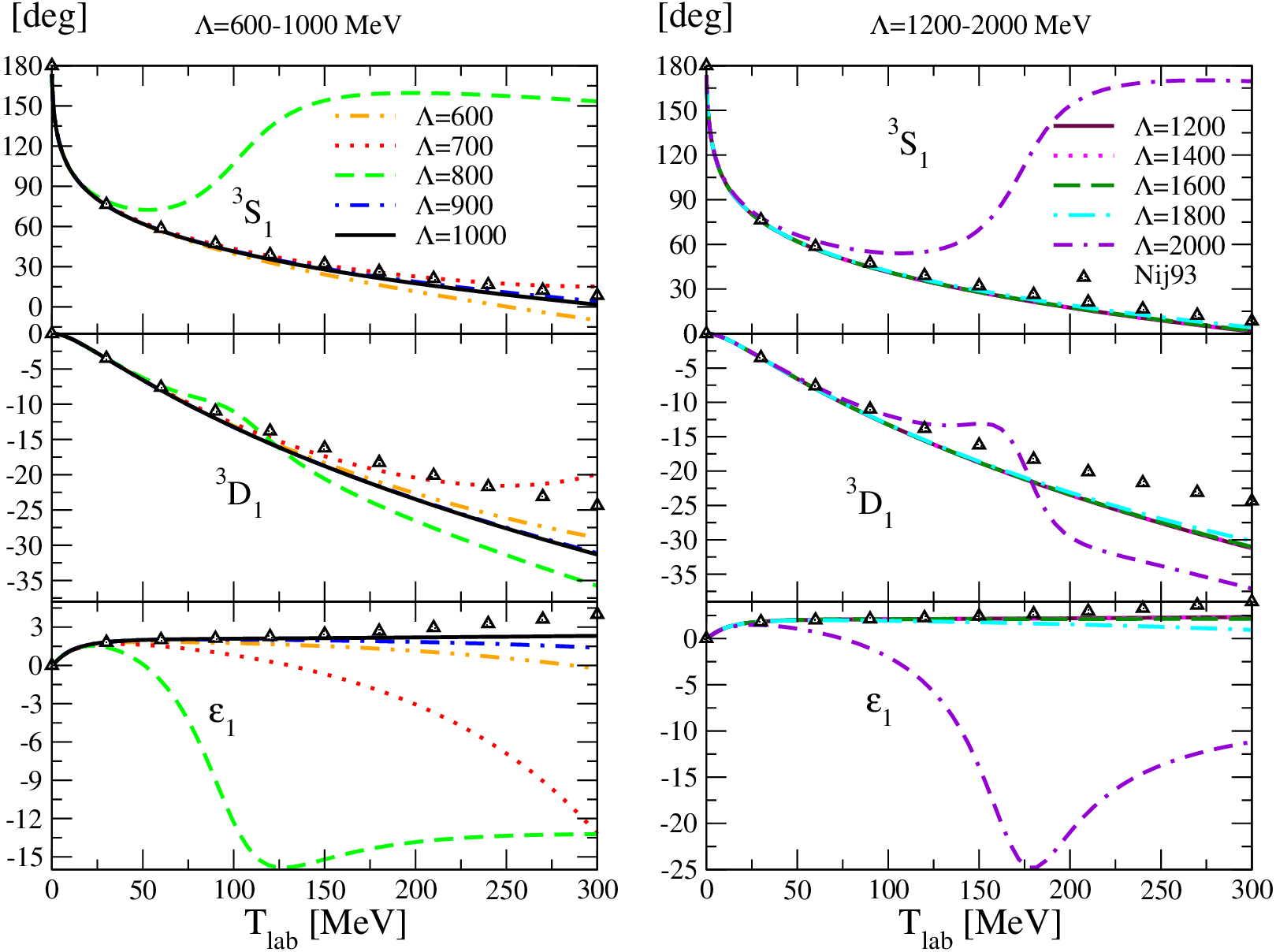}
\end{center}
\par
\vspace{3mm}
\caption{(Color online) The J=1 coupled NN phase shifts as a function of the
laboratory kinetic energy for different cutoffs $\Lambda$ ranging from 0.6
to 2~GeV. The potential employed is DR NLO with a linear energy dependence
in the central part of the contact term. The results are obtained by three
subtractions with $a_0=5.428$ fm, $\protect\alpha_{20}=2.25\times 10^{-10}$
MeV$^{-3}$ and the phase shift at $T_{lab}=10$ MeV as input. The values of
the Nijmegen phase-shifts~\protect\cite{nnonline} are indicated by the open
triangles.}
\label{fig-fig16}
\end{figure}

\begin{figure}[tbp]
\begin{center}
\includegraphics[width=16cm]{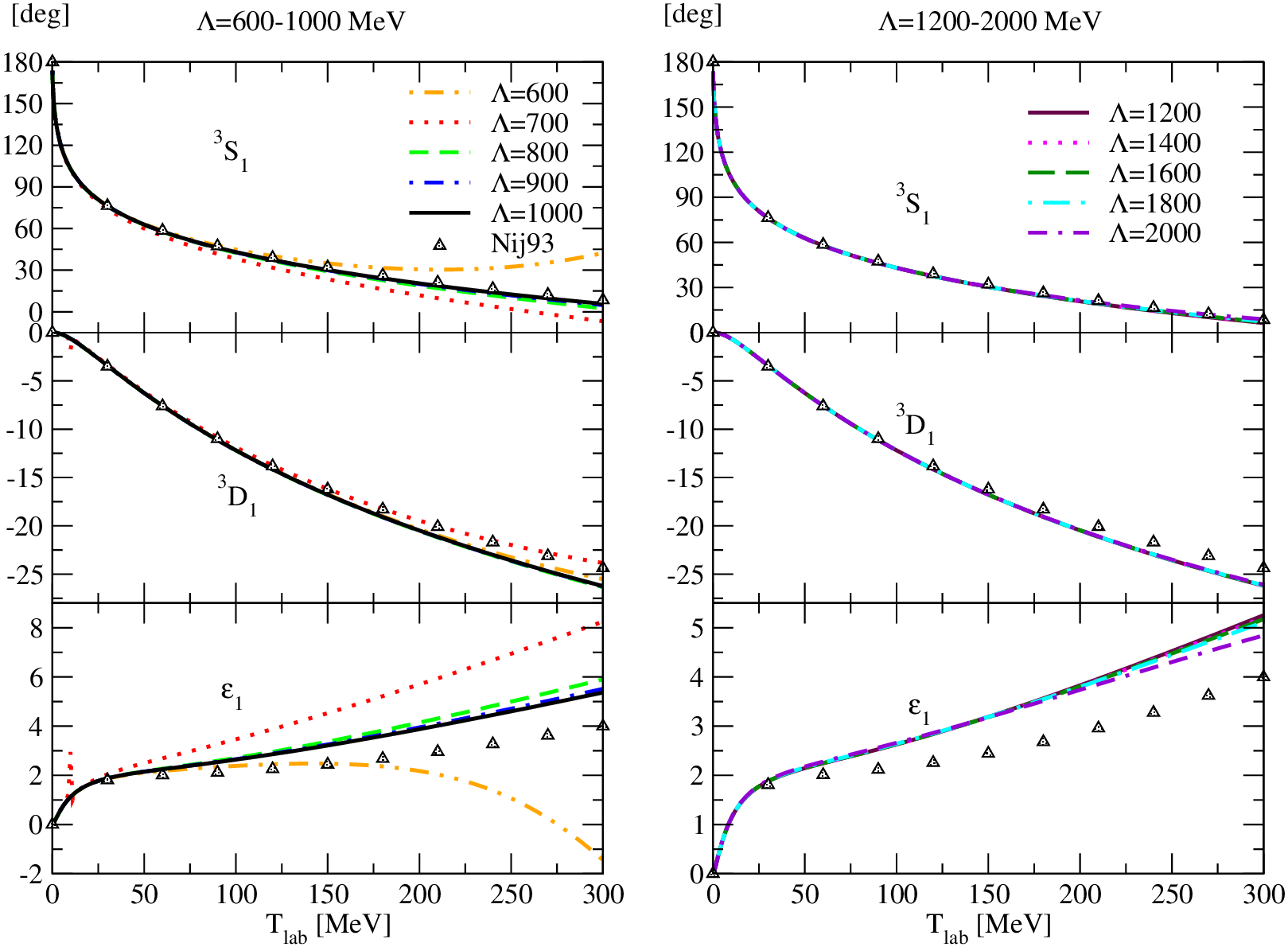}
\end{center}
\par
\vspace{3mm}
\caption{(Color online) The J=1 coupled NN phase shifts as a function of the
laboratory kinetic energy for different cutoffs $\Lambda$ ranging from 0.6
to 2~GeV. The potential employed is SFR NNLO with a linear energy dependence
in the central part of the contact term. The results are obtained by three
subtractions with $a_0=5.428$ fm, $\protect\alpha_{20}=2.25\times 10^{-10}$
MeV$^{-3}$ and the phase shift at $T_{lab}=10$ MeV as input. The values of
the Nijmegen phase-shifts~\protect\cite{nnonline} are indicated by the open
triangles.}
\label{fig-fulle}
\end{figure}

\begin{figure}[tbp]
\begin{center}
\includegraphics[width=8cm]{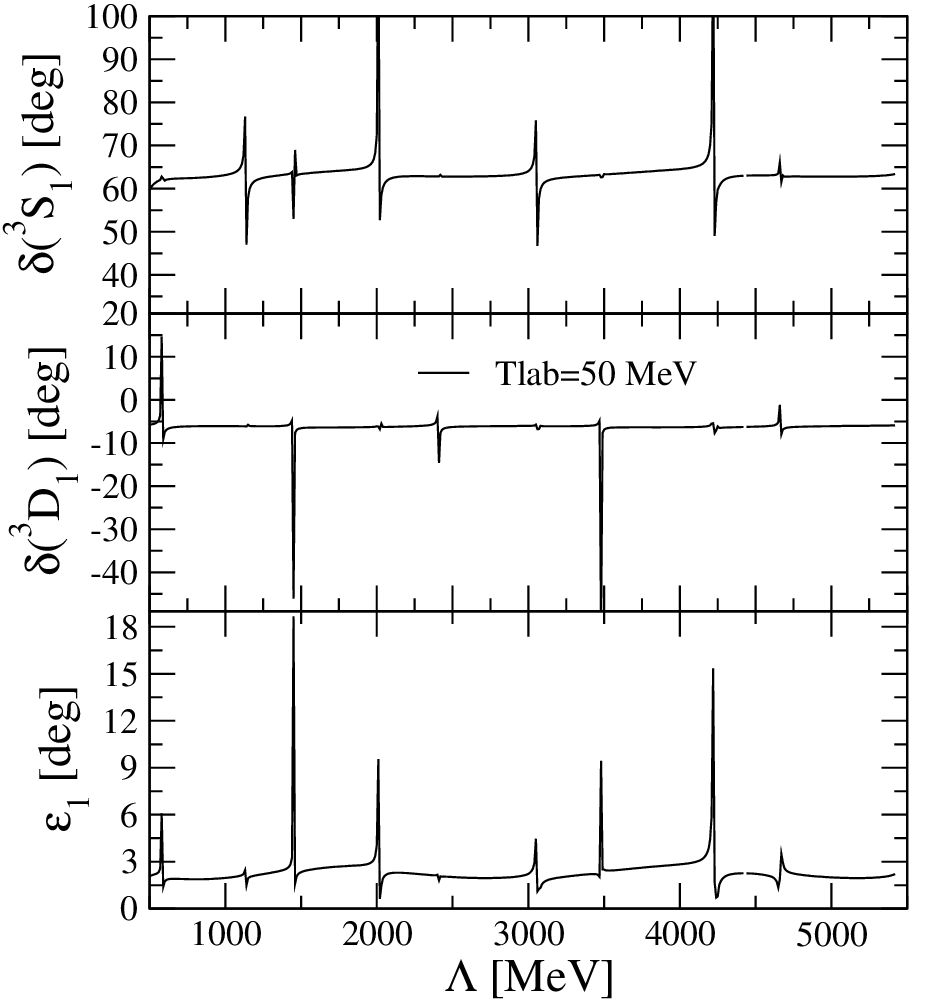}
\end{center}
\par
\vspace{3mm}
\caption{(Color online) The J=1 coupled NN phase shifts at $T_{lab}=50$~MeV
as a function of cutoff ranging from 0.5--5.5~GeV. The results are obtained
with the DR NNLO potential and a linear energy dependence in the central
part of the contact term.}
\label{fig-fig18}
\end{figure}


\begin{figure}[tbp]
\begin{center}
\includegraphics[width=8cm]{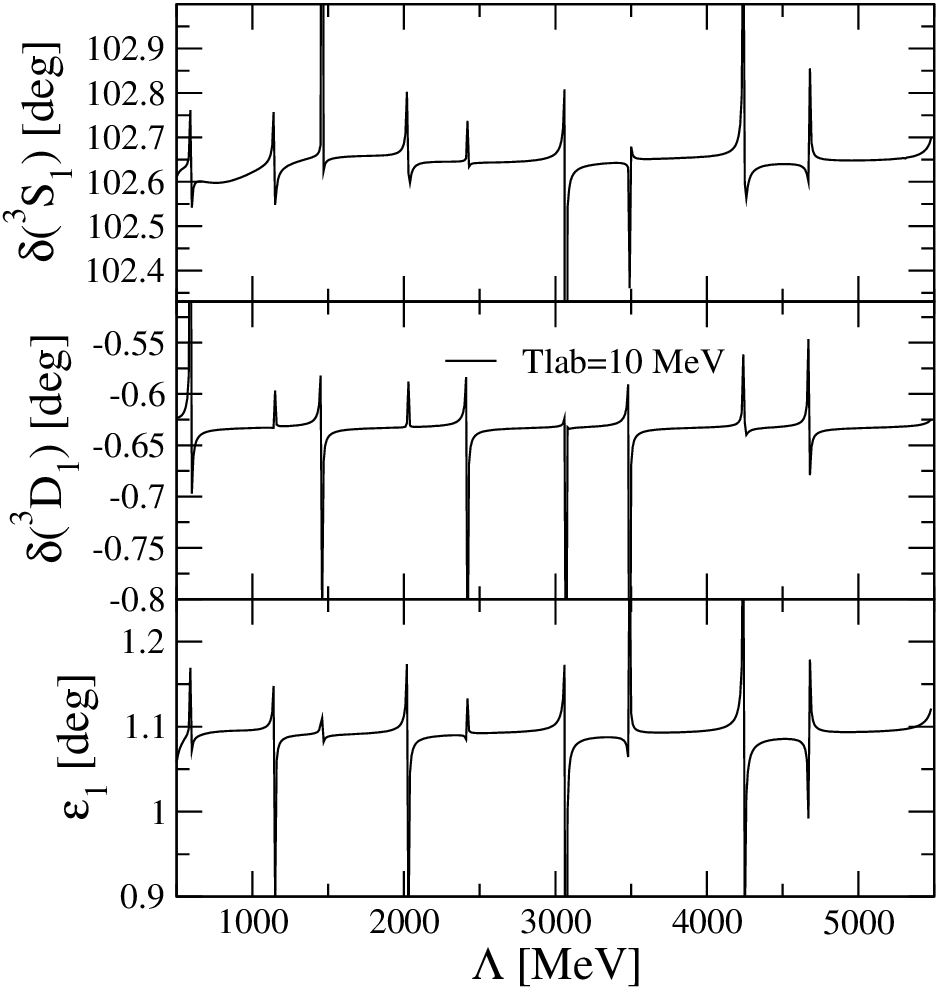}
\end{center}
\par
\vspace{3mm}
\caption{(Color online) The J=1 coupled NN phase shifts at $T_{lab}=10$~MeV
as a function of cutoff ranging from 0.5--5.5~GeV. The results are obtained
with the DR NNLO potential and a linear energy dependence in the central
part of the contact term.}
\label{fig-fig188}
\end{figure}

\begin{figure}[tbp]
\begin{center}
\includegraphics[width=16cm]{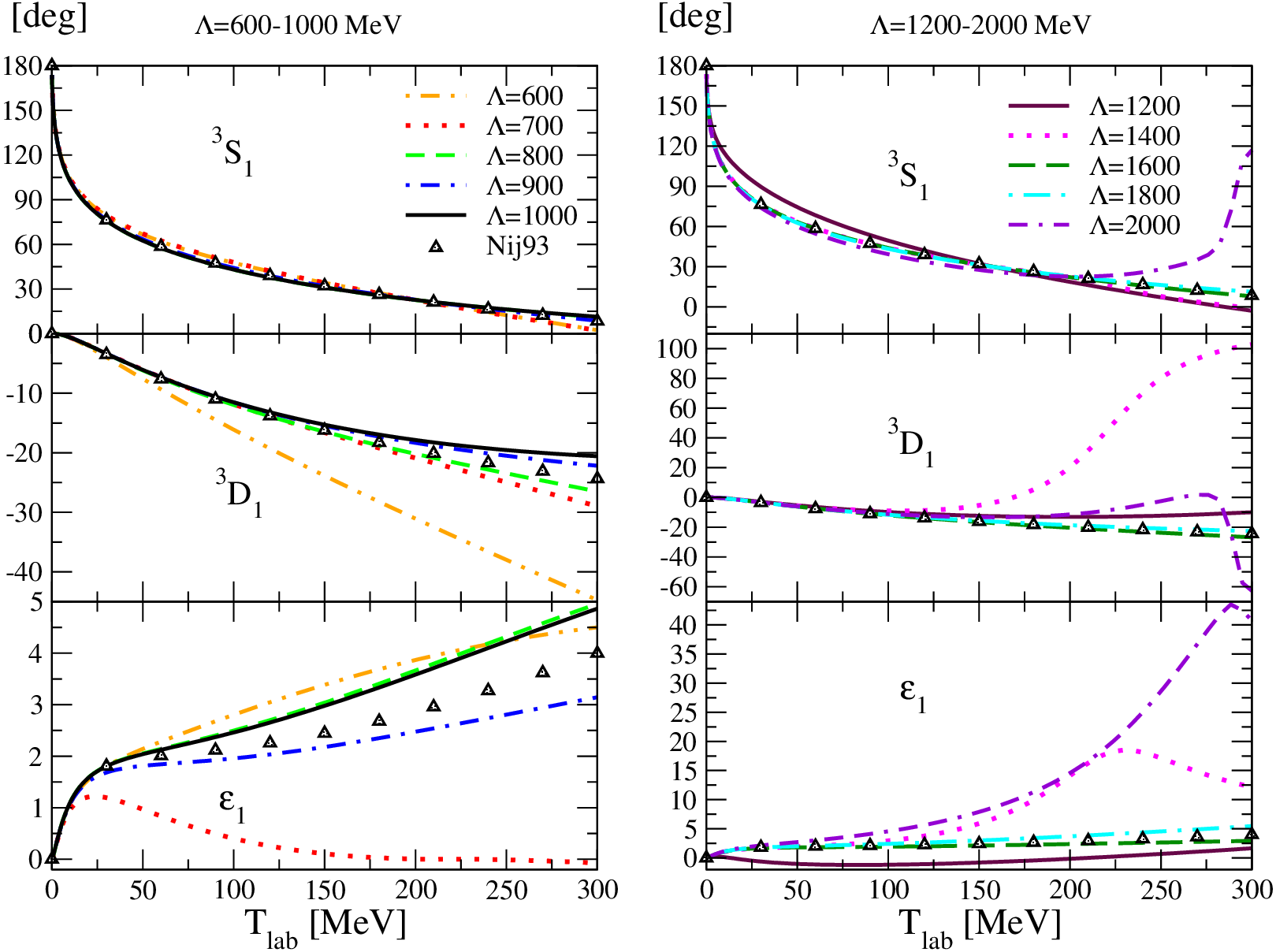}
\end{center}
\par
\vspace{3mm}
\caption{(Color online) The J=1 coupled NN phase shifts as a function of the
laboratory kinetic energy for different cutoffs $\Lambda$ ranging from 0.6
to 2~GeV. The potential employed is DR NNLO with the momentum-dependent
contact term. The results are obtained by two subtractions with $a_0=5.428$
fm and $\protect\alpha_{20}=2.25\times 10^{-10}$ MeV$^{-3}$ as input and
then performing a fit to the $^3S_1$ Nijmegen phase shift at $T_{lab}=200$%
~MeV. The values of the Nijmegen phase-shifts~\protect\cite{nnonline} are
indicated by the open triangles.}
\label{fig-fig19}
\end{figure}

\begin{figure}[tbp]
\begin{center}
\includegraphics[width=16cm]{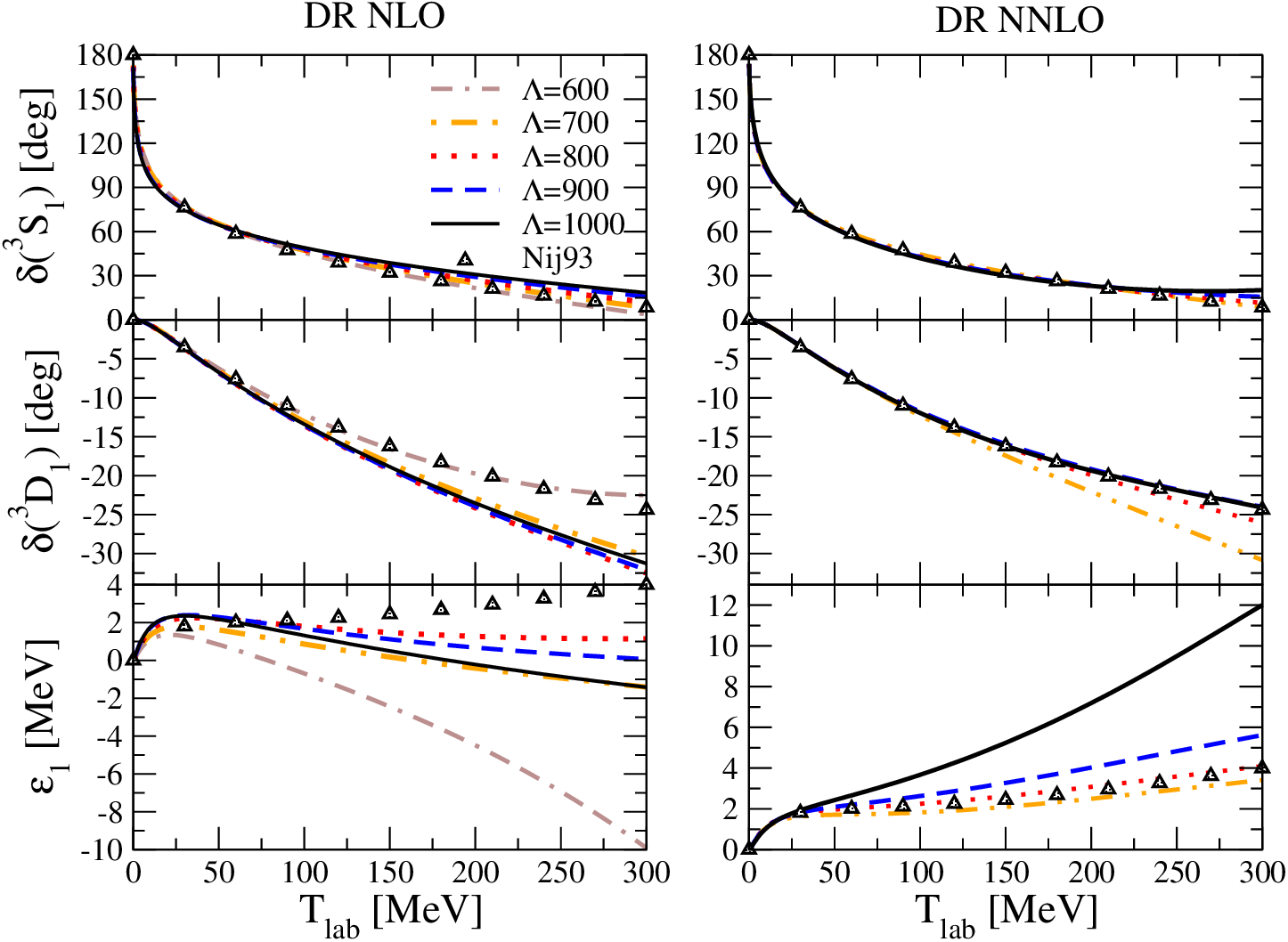}
\end{center}
\par
\vspace{3mm}
\caption{(Color online) The best fit for the NN $^3$S$_1-^3$D$_1$ phase
shifts as a function of the laboratory kinetic energy for different cutoffs $%
\Lambda$ ranging from 0.6 to 1~GeV. The potentials employed are the DR NLO
(left panel) and the DR NNLO (right panel) with a momentum-dependent central
part of the contact term. The values of the Nijmegen phase-shifts~%
\protect\cite{nnonline} are indicated by the open triangles. }
\label{fig-fig21a}
\end{figure}

\begin{figure}[tbp]
\begin{center}
\includegraphics[width=8cm]{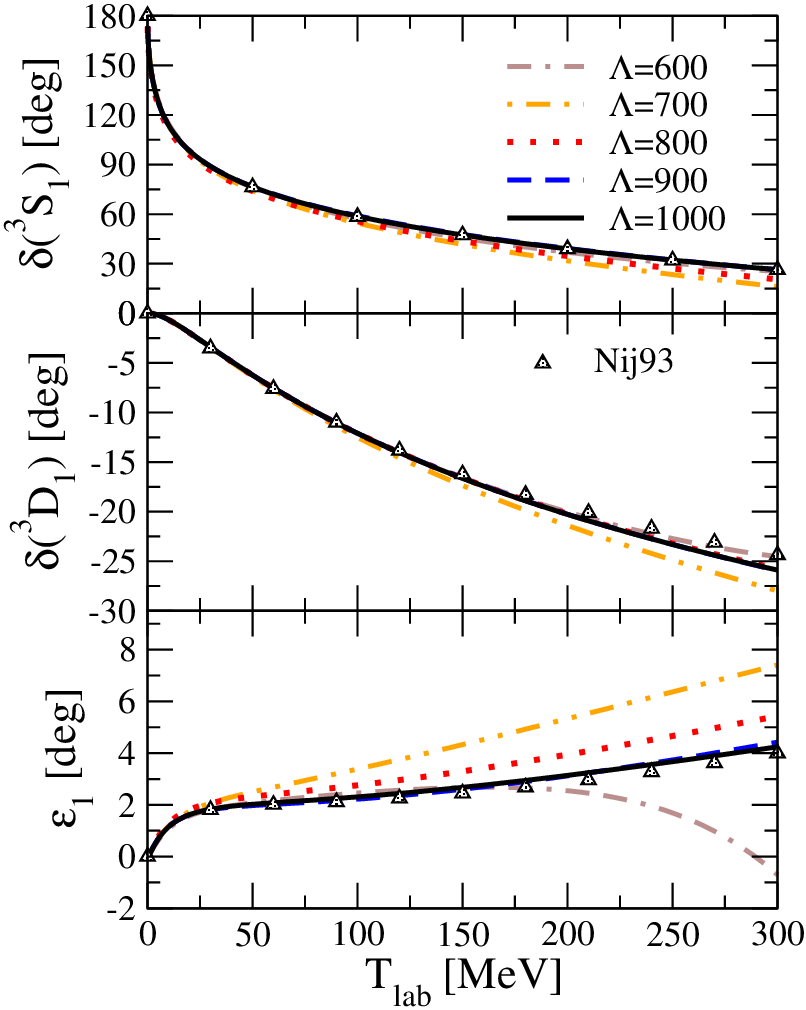}
\end{center}
\par
\vspace{3mm}
\caption{(Color online) The best fit for the NN $^3$S$_1-^3$D$_1$ phase
shifts as a function of the laboratory kinetic energy for different cutoffs $%
\Lambda$ ranging from 0.6 to 1~GeV. The potentials employed are the SFR NNLO
with a momentum-dependent central part of the contact term. The values of
the Nijmegen phase-shifts~\protect\cite{nnonline} are indicated by the open
triangles. }
\label{fig-fig200}
\end{figure}

\begin{figure}[tbp]
\begin{center}
\includegraphics[width=16cm]{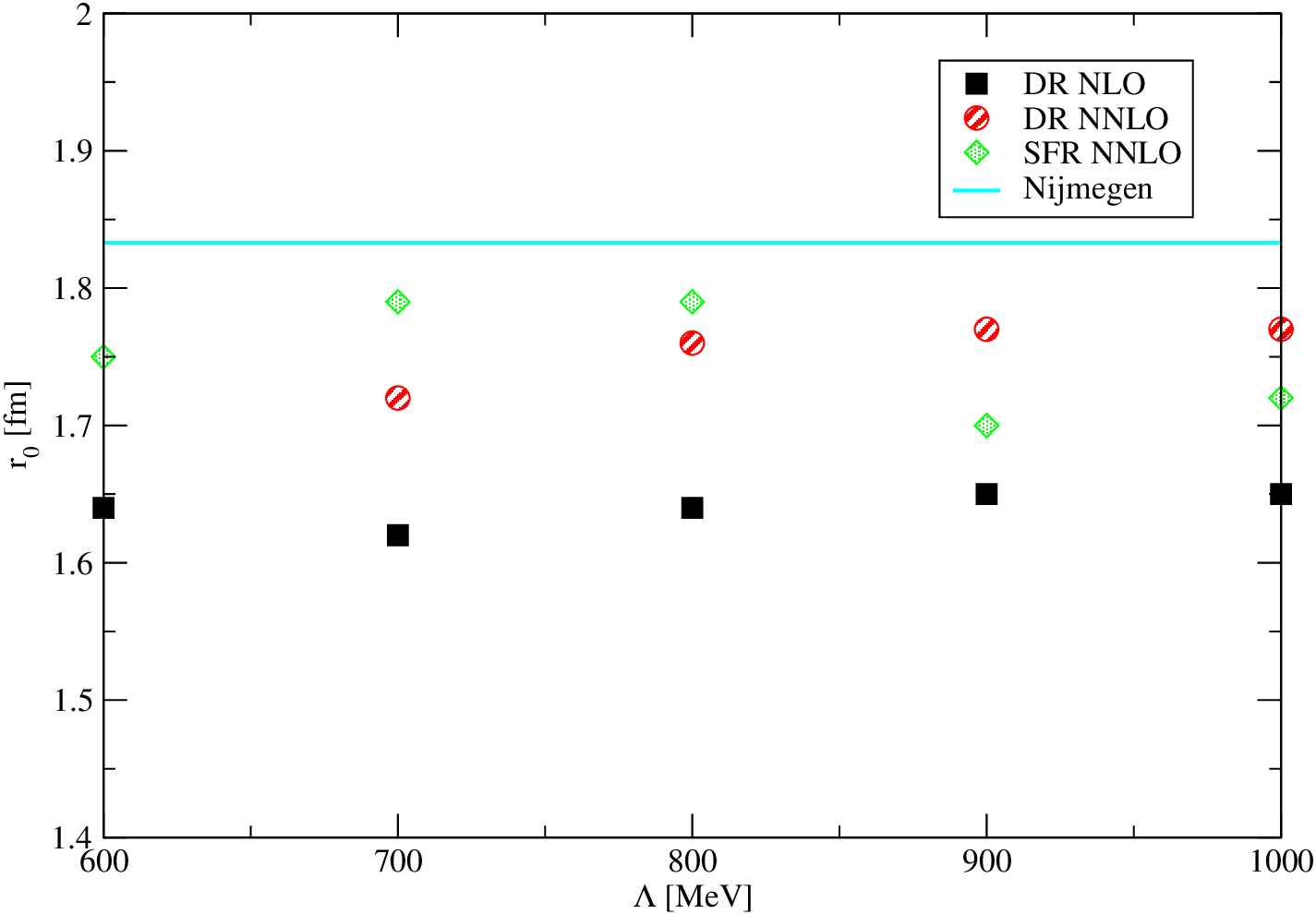}
\end{center}
\par
\vspace{3mm}
\caption{The effective range $r_{0}$ [in fm] in the $^3$S$_1$ channel
extracted from calculations with the DR~NLO (black square), DR~NNLO (red circle) and SFR~NNLO (green diamond) TPE
 and a momentum-dependent central piece of the contact term. Here $r_0$ is
shown as a function of the cutoff $\Lambda$ in the LSE, and is extracted
from a best fit of the phase shifts to the Nijmegen PWA93. The solid line represents $r_0$ obtained from Ref.\cite{PVRA05B}. Note that the value of $r_0$ at $\Lambda=600$ MeV for DR NNLO is $-11.5$ fm, which is not plotted in the figure.}
\label{fig-t4}
\end{figure}

\end{document}